\documentclass[prb,aps,twocolumn,footinbib,superscriptaddress,floatfix,bibliography]{revtex4-1}
\usepackage{CJK}
\usepackage[colorlinks,bookmarks=false,citecolor=blue,linkcolor=red,urlcolor=blue]{hyperref}
\usepackage{color} 
\usepackage{graphicx}
\usepackage{float}
\usepackage{multirow}
\usepackage{amsmath}
\usepackage{bm}
\usepackage{amsfonts}
\usepackage{braket}
\usepackage{subfigure}
\usepackage{cleveref}

\begin{document}
%%%%%%%%%%%%%%%%%%%%%%%%%%%%%%%%%%%%%%%%%%%%%%%
\title{Ground state phase diagram of dipolar-octupolar pyrochlores}

\author{Owen Benton}
\affiliation{RIKEN Center for Emergent Matter Science (CEMS), Wako, Saitama, 351-0198, Japan}
\affiliation{Max Planck Institute for the Physics of Complex Systems, N{\"o}thnitzer Str. 38, Dresden 01187, Germany}

\begin{abstract}
The  ``dipolar-octupolar” pyrochlore oxides R$_2$M$_2$O$_7$ (R=Ce, Sm, Nd) represent an important opportunity in 
the search for three dimensional Quantum Spin Liquid (QSL) ground states.
Their low energy physics is governed by an alluringly simple ``XYZ'' Hamiltonian, enabling theoretical description with only a small number of free parameters.
Meanwhile, recent experiments on Ce pyrochlores strongly suggest QSL physics.
Motivated by this, we present here a complete analysis of the ground state 
phase diagram of dipolar-octupolar pyrochlores. 
Combining cluster mean field theory, 
variational arguments and exact diagonalization we find multiple U(1) QSL phases which 
together occupy a large fraction of the parameter space. 
These results give a comprehensive picture of the ground state physics of an important class of QSL candidates and
support the possibility of a $U(1)$ QSL ground state in Ce$_2$Zr$_2$O$_7$ and  Ce$_2$Sn$_2$O$_7$.
\end{abstract}

\maketitle

\section{Introduction}
\label{sec:intro}

The pursuit of quantum spin liquid (QSL) ground states
has not gone unrewarded.
On the theory side, it has been realized that an enormous diversity of
QSL states are possible \cite{wen02, Savary2016} and several physically relevant 
models are now known
to have QSL ground states 
\cite{kitaev06-AnnPhys321, hermele04PRB, banerjee08, Gingras14RoPP, lee14, hu15, iqbal16, deppenbrock12, he17}.
In experiment, many candidate materials have been established,
exhibiting spin liquid like properties at low temperature 
\cite{okamoto07, Banerjee2016, balz16-NatPhys12, norman16, sibille18}.

What has yet to be achieved is the combination of a material with an experimentally robust QSL state,
with theoretical understanding of the microscopic interactions which
give rise to that state and what kind of spin liquid they produce.
Some materials studied as potential QSLs actually order at low temperature
\cite{chang12, cao16, takatsu16}, and others are complicated
by chemical or structural disorder \cite{kermarrec14, han16, martin17, zhu17}.
Meanwhile, the relevant theoretical models are often complicated, possessing many free
parameters \cite{ross11, li16, Essafi2017}.

``Dipolar-octupolar'' (DO) pyrochlores R$_2$M$_2$O$_7$ (R=Ce, Sm, Nd; M=Zr, Hf, Ti, Sn, Pb) 
\cite{sibille15, sibille-arXiv,
gao19, gaudet19, 
pecanha19, mauws18, singh08, malkin10, xu-thesis,
lhotel15, xu15, petit16-NatPhys12, 
xu19, 
anand17, bertin15,
dalmas17, 
hallas15, swarnarkar17}
constitute an
opportunity in this context,
with their low energy physics being described by a simple
XYZ Hamiltonian \cite{huang14, rau19}.
Out of this family, Ce$_2$Sn$_2$O$_7$ \cite{sibille15, sibille-arXiv} and Ce$_2$Zr$_2$O$_7$ \cite{gao19, gaudet19}
have been highlighted recently as showing evidence of QSL physics.
Notably, neutron scattering results for Ce$_2$Zr$_2$O$_7$ bear encouraging similarity to predictions for a
$U(1)$ quantum spin liquid \cite{gaudet19, benton12-PRB86}.

\begin{figure*}
\includegraphics[width=0.9\textwidth]{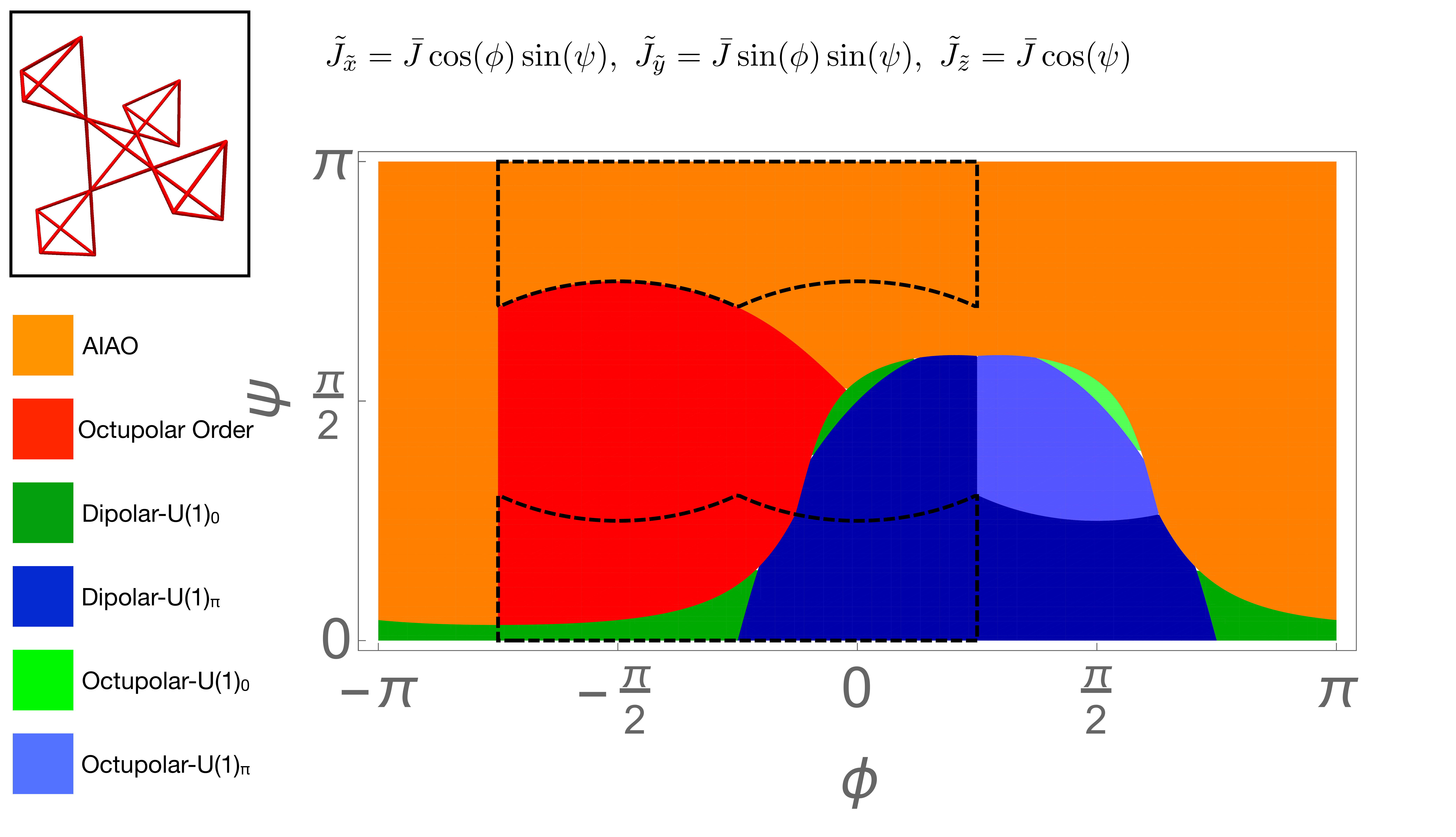}
\caption{Ground state phase diagram of the XYZ model [Eq. (\ref{eq:HXYZ})] on the pyrochlore lattice (inset), describing dipolar-octupolar pyrochlores. 
The three exchange parameters $\tilde{J}_{\tilde{\alpha}}$ are represented in terms of an overall scale ${\bar J}$ and two angular variables $\phi, \psi$
[Eq. (\ref{eq:angular_rep})].
The phase diagram features ``all in/all out'' (AIAO) and octupolar ordered phases, and four distinct U(1) QSLs.
These four QSLs are distinguished by whether the emergent electric field of the low energy
gauge theory transforms like a magnetic dipole or octupole, and by the flux penetrating elementary plaquettes in the ground state (0 or $\pi$).
The phase diagram is obtained by combining Cluster Mean Field Theory (CMFT), 
a cluster variational  (CVAR) calculation and 
Exact Diagonalization (ED) as described in the text.
The two regions bounded by black dashed lines correspond to the subset of parameters in Eq. (\ref{eq:subregion}),
from which the entire phase diagram can be generated using unitary transformations.
}
\label{fig:pd-full}
\end{figure*}

In DO pyrochlores, the magnetic rare earth ions form a corner-sharing tetrahedral structure [Fig.
\ref{fig:pd-full} (inset)].
There are strong crystal electric fields (CEFs) acting on each magnetic site, resulting in a Kramers
doublet at the bottom of the CEF spectrum, separated from higher states
by a large gap $\Delta_{CEF}\sim100{\rm K}$ \cite{xu15, anand17, gaudet19}.
With the scale of exchange interactions being $\sim1 {\rm K}$ \cite{sibille-arXiv, xu19}, this motivates a description of
the system in terms of pseudospin-1/2 operators $\tau^x_i, \tau^y_i, \tau^z_i$.
The thing which sets DO pyrochlores apart from other pyrochlore oxides is the transformation
properties of these operators under time-reversal and lattice symmetries \cite{huang14, rau19}.
 $\tau^x_i$ and $\tau^z_i$ both transform like the component of a magnetic dipole oriented along
 the site's $C_3$ symmetry axis, while $\tau^y_i$ transforms like a component of the magnetic octupole
 tensor.
 
 Assuming nearest-neighbor  interactions, symmetry constrains the Hamiltonian
  to take the  form \cite{huang14}:
 \begin{eqnarray}
&&\mathcal{H}=\sum_{\langle ij \rangle} \left[ 
\left(\sum_{\alpha=x,y,z}
{J}_{\alpha} \tau^{\alpha}_i  \tau^{\alpha}_j\right)
+ J_{xz} \left(
 \tau^x_i  \tau^z_j
+ \tau^z_i  \tau^x_j
\right)
\right].
\label{eq:H0}
 \end{eqnarray}
The final term in Eq. (\ref{eq:H0}) 
can be removed by a suitably chosen global transformation $\tau^{\alpha} \to \tilde{\tau}^{\tilde{\alpha}}$
\cite{huang14, Benton2016PRB},
reducing the problem to an XYZ Hamiltonian:
\begin{eqnarray}
\mathcal{H}=\sum_{\langle ij \rangle} 
\sum_{\alpha=\tilde{x},\tilde{y},\tilde{z}}
\tilde{J}_{\alpha} \tilde{\tau}^{\alpha}_i  \tilde{\tau}^{\alpha}_j.
\label{eq:HXYZ}
 \end{eqnarray}
 
 An understanding of dipolar-octupolar pyrochlores and their potential to realize QSL
 ground states requires understanding of the ground state phase
 diagram of Eq. (\ref{eq:HXYZ}).
Certain limits of the parameter space of Eq. (\ref{eq:HXYZ}) have been well studied, 
namely: the perturbative limit where one exchange parameter dominates the other two 
\cite{hermele04PRB, shannon12, benton12-PRB86},
the XXZ limit where two of the three exchange parameters are equal 
\cite{savary12, lee12, benton18-PRL121, banerjee08, kato15, huang18}
and
the region of parameter space without a sign problem for Quantum Monte Carlo (QMC) 
\cite{huang14, banerjee08, kato15, huang18, huang-arxiv}. 
However,  there is no reason to expect materials of interest to fall into
one of these limits, so a global phase diagram is needed.

 In this Article, we calculate the ground state phase diagram of  Eq. (\ref{eq:HXYZ}),
by combining Cluster Mean Field Theory (CMFT), 
 a variational extension to CMFT (CVAR) \cite{benton18-PRL121} and 
 Exact Diagonalization (ED).
 Where the results can be compared with available QMC results \cite{huang-arxiv}, they agree well.
 The final result for the phase diagram is shown in Fig. \ref{fig:pd-full}, with the parameter
 space expressed in terms of an overall scale ${\bar J}$ which can be divided out and
 two angles $\phi, \psi$:
 \begin{eqnarray}
 &&\tilde{J}_{\tilde{x}}={\bar J} \cos(\phi) \sin(\psi),\
 \tilde{J}_{\tilde{y}}={\bar J} \sin(\phi) \sin(\psi),\ \nonumber \\
&&  \tilde{J}_{\tilde{z}}={\bar J} \cos(\psi)
\label{eq:angular_rep}
 \end{eqnarray}
 
 We find four $U(1)$ spin liquid phases, occupying a large combined portion 
 of the parameter space, competing with an antiferromagnetic ``all in/all out''
 (AIAO) phase and octupolar order. 
 The four $U(1)$ QSLs all host gapless photons and gapped fractionalized charges,
 and are thus realizations of emergent electromagnetism \cite{hermele04PRB, shannon12, 
 benton12-PRB86, Gingras14RoPP}.
 They are labelled  dipolar/octupolar-$U(1)_{0/\pi}$ with the dipolar/octupolar label referring to whether the emergent electric field transforms like a magnetic dipole or octupole \cite{li17,li19}, and the $0/\pi$ subscript referring to the $U(1)$ flux penetrating elementary plaquettes in the ground state.

The remainder of this Article is devoted to explaining the calculations leading to Fig.  \ref{fig:pd-full},
before finishing with a brief discussion of the outlook for experiments.

The Article is structured as follows:
\begin{itemize}
\item{In Section \ref{sec:dualities} we describe some simple dualities which allow the whole
phase diagram to be generated from calculations covering only a subregion of parameter space.}
\item{In Section \ref{sec:CMFT} we calculate the ground state phase diagram using CMFT, augmented with the CVAR approach.}
\item{In Section \ref{sec:ED} we show ED calculations on a 16-site cluster, and use these as an alternative
route to calculate the ground state phase diagram.}
\item{The construction of the complete phase diagram [Fig.  \ref{fig:pd-full}], from the combination of the calculations in the preceding sections, is then described in Section~\ref{sec:fullpd}}.
\item{Section \ref{sec:summary} gives a summary of the results and an outlook for future work on dipolar-octupolar pyrochlores.}
\end{itemize}

\section{Dualities of the model and reduced parameter space}
\label{sec:dualities}

In calculating the phase diagram it is useful to note that Eq. (\ref{eq:HXYZ}) has some
dualities in which the exchange parameters can be permuted by a unitary transformation
acting on $\mathcal{H}$. 
Specifically:
\begin{eqnarray}
&&\mathcal{H}(\tilde{J}_{\tilde{z}}, \tilde{J}_{\tilde{x}}, \tilde{J}_{\tilde{y}})=
\mathcal{U}_{2\pi/3, 111} \mathcal{H}(\tilde{J}_{\tilde{x}}, \tilde{J}_{\tilde{y}}, \tilde{J}_{\tilde{z}}) \mathcal{U}_{2\pi/3, 111}^{\dagger} \nonumber \\
&&\mathcal{H}(\tilde{J}_{\tilde{y}}, \tilde{J}_{\tilde{x}}, \tilde{J}_{\tilde{z}})=
\mathcal{U}_{\pi/2, 001} \mathcal{H}(\tilde{J}_{\tilde{x}}, \tilde{J}_{\tilde{y}}, \tilde{J}_{\tilde{z}}) \mathcal{U}_{\pi/2, 001}^{\dagger}
\label{eq:dualities}
\end{eqnarray}
where $\mathcal{U}_{\gamma, {\bf v}}$ represents a global rotation by an angle $\gamma$ around axis ${\bf v}$ of pseudospin space (which is not the same as a rotation in the physical crystal space).
Making use of these dualities means that we do not actually need to study the full parameter space of $\tilde{J}_{\tilde{x}}, \tilde{J}_{\tilde{y}}, \tilde{J}_{\tilde{z}}$, it is enough to consider a subset of parameters
\begin{eqnarray}
|\tilde{J}_{\tilde{z}}| > |\tilde{J}_{\tilde{x}}|, |\tilde{J}_{\tilde{y}}|, \quad \tilde{J}_{\tilde{x}}>  \tilde{J}_{\tilde{y}}
\label{eq:subregion}
\end{eqnarray}
from which we can then generate the rest of the phase diagram by applying the transformations from Eq. (\ref{eq:dualities}) to our results.
The parameter space dsecribed by (\ref{eq:subregion}) is delineated by the black dashed lines in Fig. \ref{fig:pd-full}.
\begin{figure}
\centering
\includegraphics[width=0.4\textwidth]{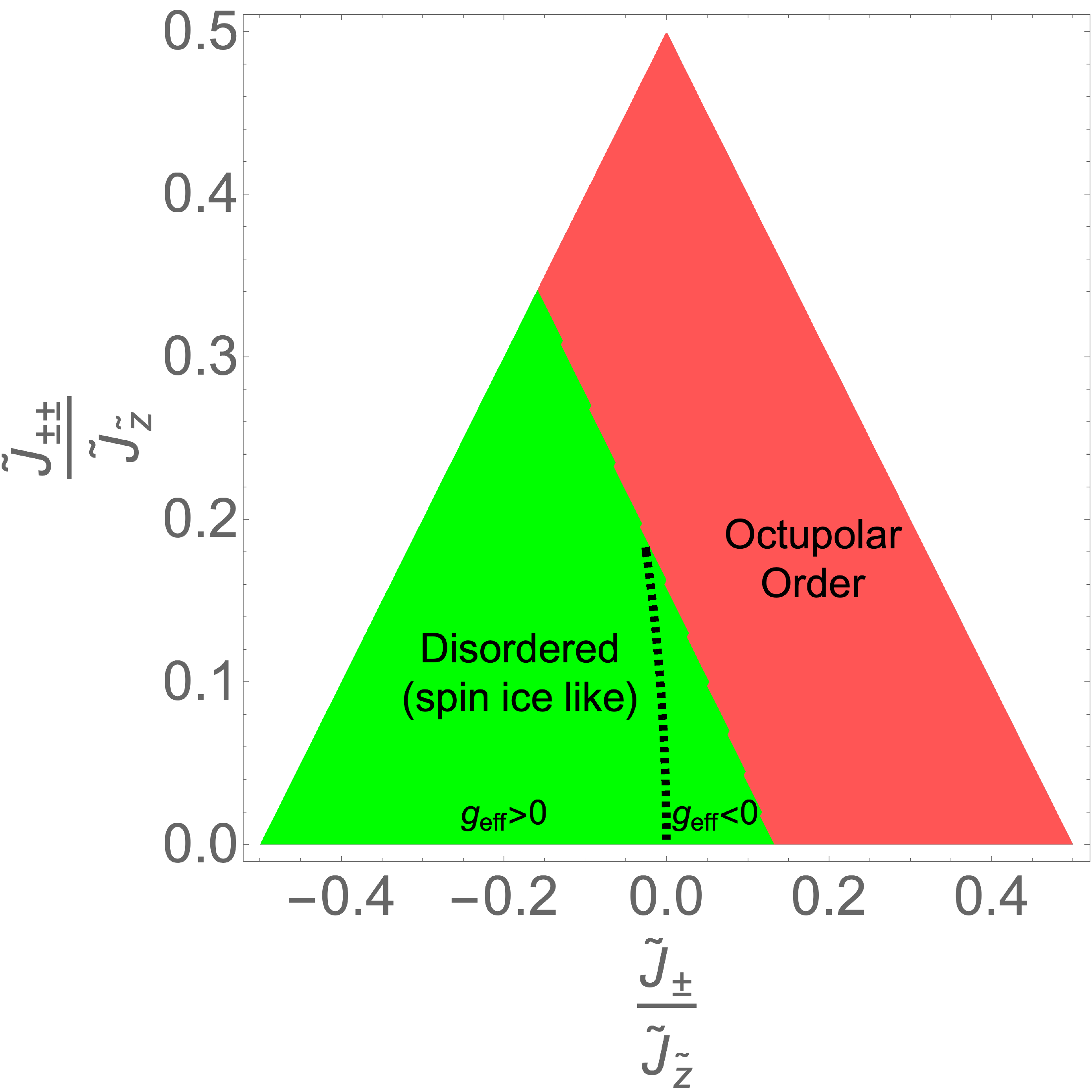}
\centering
\caption{CMFT  and CVAR calculations of the ground state phase diagram of $\mathcal{H}$
[Eqs. (\ref{eq:HXYZ}), (\ref{eq:Hpm})] within 
the region of parameter space given by (\ref{eq:subregionpm}) with $\tilde{J}_{\tilde{z}}>0$.
CMFT calculations give two regimes for the optimal configuration of the auxiliary fields ${\bf h}_i$:
an ordered region where the ${\bf h}_i$ point uniformly along the $y$ axis of pseudospin space (red) and
a disordered region with a large degeneracy of CMFT solutions where  ${\bf h}_i=\sigma_i h \tilde{{\bf z}}_i$ ,
with signs $\sigma_i$ summing to zero on every tetrahedron (green).
The CVAR calculation, which incorporates quantum tunnelling between CMFT solutions,
breaks the degenerate region into two, based on the sign of the 
effective tunnelling matrix element $g_{eff}$. 
Positive (negative) values of $g_{eff}$ lead ultimately to a $\pi$-flux (0-flux) U(1) QSL ground state.}
\label{fig:numerics}
\end{figure}

Taking $\tilde{J}_{\tilde{z}}$ to be the strongest exchange parameter as in (\ref{eq:subregion}), if $\tilde{J}_{\tilde{z}}<0$  it is clear that the ground state
will simply order ferromagnetically with respect to the  $\tilde{z}$-axis of pseudospin space.
In terms of the physical magnetic moments this implies AIAO order.
The more challenging problem is to discover what happens when $\tilde{J}_{\tilde{z}}>0$.

To study this case we rewrite the Hamiltonian in terms of spin ladder operators
$\tilde{\tau}^{{\pm}}_i$:
\begin{eqnarray}
&&\mathcal{H}=\sum_{\langle ij\rangle} \bigg[ \tilde{J}_{\tilde{z}} \tilde{\tau}^{\tilde{z}}_i
 \tilde{\tau}^{\tilde{z}}_j -
 \tilde{J}_{\pm} \left(
 \tilde{\tau}^{{+}}_i  \tilde{\tau}^{{-}}_j+ \tilde{\tau}^{{-}}_i  \tilde{\tau}^{{+}}_j
 \right)
 \nonumber \\
&&\qquad+
 \tilde{J}_{\pm\pm} \left(
 \tilde{\tau}^{{+}}_i  \tilde{\tau}^{{+}}_j+ \tilde{\tau}^{{-}}_i  \tilde{\tau}^{{-}}_j
 \right)
 \bigg]
 \label{eq:Hpm}
\end{eqnarray}
where
$\tilde{J}_{\pm}=-\frac{1}{4} \left(\tilde{J}_{\tilde{x}} +\tilde{J}_{\tilde{y}}   \right)$
and 
$\tilde{J}_{\pm\pm}=\frac{1}{4} \left(\tilde{J}_{\tilde{x}} -\tilde{J}_{\tilde{y}}   \right)$.
The subregion of parameter space given by (\ref{eq:subregion}) then becomes:
\begin{eqnarray}
|2(\tilde{J}_{\pm\pm}-\tilde{J}_{\pm})|< \tilde{J}_{\tilde{z}}, \
|2(\tilde{J}_{\pm\pm}+\tilde{J}_{\pm})|<\tilde{J}_{\tilde{z}}, \
\tilde{J}_{\pm\pm}>0. \ \
\label{eq:subregionpm}
\end{eqnarray}

\section{Phase diagram from Cluster Mean Field Theory}
\label{sec:CMFT}

\subsection{CMFT Calculation}
\label{subsec:CMFT}

To begin, we consider the phase diagram using a tetrahedral CMFT, as employed for
the XXZ limit ($\tilde{J}_{\pm\pm}=0$) in Ref. \onlinecite{benton18-PRL121}.
A summary of the calculation is given here, with a detailed description found in Appendix \ref{sec:CMFTapp}.

To construct the CMFT we use the fact that the pyrochlore lattice can be 
divided into two sets of tetrahedra `A' and `B', with all neighbors of an `A' tetrahedron
being `B' tetrahedra and vice versa.
We then seek to optimize a product wave function over all 
`A' tetrahedra:
\begin{eqnarray}
|\psi_{\sf CMFT} \rangle=\prod_{t \in A} |\phi_t \rangle.
\end{eqnarray}
The wave function $|\phi_t\rangle$ on each tetrahedron $t$ is defined to be
the ground state of a single tetrahedron Hamiltonian 
\begin{eqnarray}
\mathcal{H}'_t |\phi_t \rangle= \epsilon_{0, t} |\phi_t \rangle.
\end{eqnarray}
$\mathcal{H}'_t$
contains the original exchange terms acting on the bonds of $t$ as well as auxiliary
fields ${\bf h}_i$ on each site
\begin{eqnarray}
&&\mathcal{H}'_t=\sum_{\langle ij \rangle \in t} \bigg[ \tilde{J}_{\tilde{z}} \tilde{\tau}^{\tilde{z}}_i
 \tilde{\tau}^{\tilde{z}}_j -
 \tilde{J}_{\pm} \left(
 \tilde{\tau}^{{+}}_i  \tilde{\tau}^{{-}}_j+ \tilde{\tau}^{{-}}_i  \tilde{\tau}^{{+}}_j
 \right)
 \nonumber \\
&&\qquad+
 \tilde{J}_{\pm\pm} \left(
 \tilde{\tau}^{{+}}_i  \tilde{\tau}^{{+}}_j+ \tilde{\tau}^{{-}}_i  \tilde{\tau}^{{-}}_j
 \right)
 \bigg] - \sum_{i \in t} \sum_{\alpha =\tilde{x}, \tilde{y}, \tilde{z}}h^{\alpha}_i  \tilde{\tau}^{\alpha}_i.
\end{eqnarray}
The auxiliary fields ${\bf h}_i$ then serve as variational parameters for optimizing
$|\psi_{\sf CMFT} \rangle$, and a CMFT wave function can be indexed by
a configuration of  ${\bf h}_i$ on the lattice.

There are two regimes for the optimal configuration of ${\bf h}_i$ in CMFT as shown in 
Fig. \ref{fig:numerics}.
For sufficiently large, positive, values of $\tilde{J}_{\pm}$ or $\tilde{J}_{\pm\pm}$ the optimal
solutions have  ${\bf h}_i$ ordered ferromagnetically along the $y$-axis of pseudospin space.
This implies $\langle \tilde{\tau}^{\tilde{y}} \rangle \neq0$, and therefore octupolar order
since  $\tilde{\tau}^{\tilde{y}}$ transforms like a magnetic octupole \cite{huang14}.

In the remainder of the phase diagram there is a large, ice-like, degeneracy of disordered CMFT solutions,
with ${\bf h}_i=\sigma_i h \tilde{{\bf z}}_i$ where $h$ is a fixed, uniform, magnitude and 
$\sigma_i=\pm1$, subject to the constraint that $\sigma_i$ sum to zero on every tetrahedron.

\begin{figure*}
\centering
\subfigure[]{\includegraphics[width=0.4\textwidth]{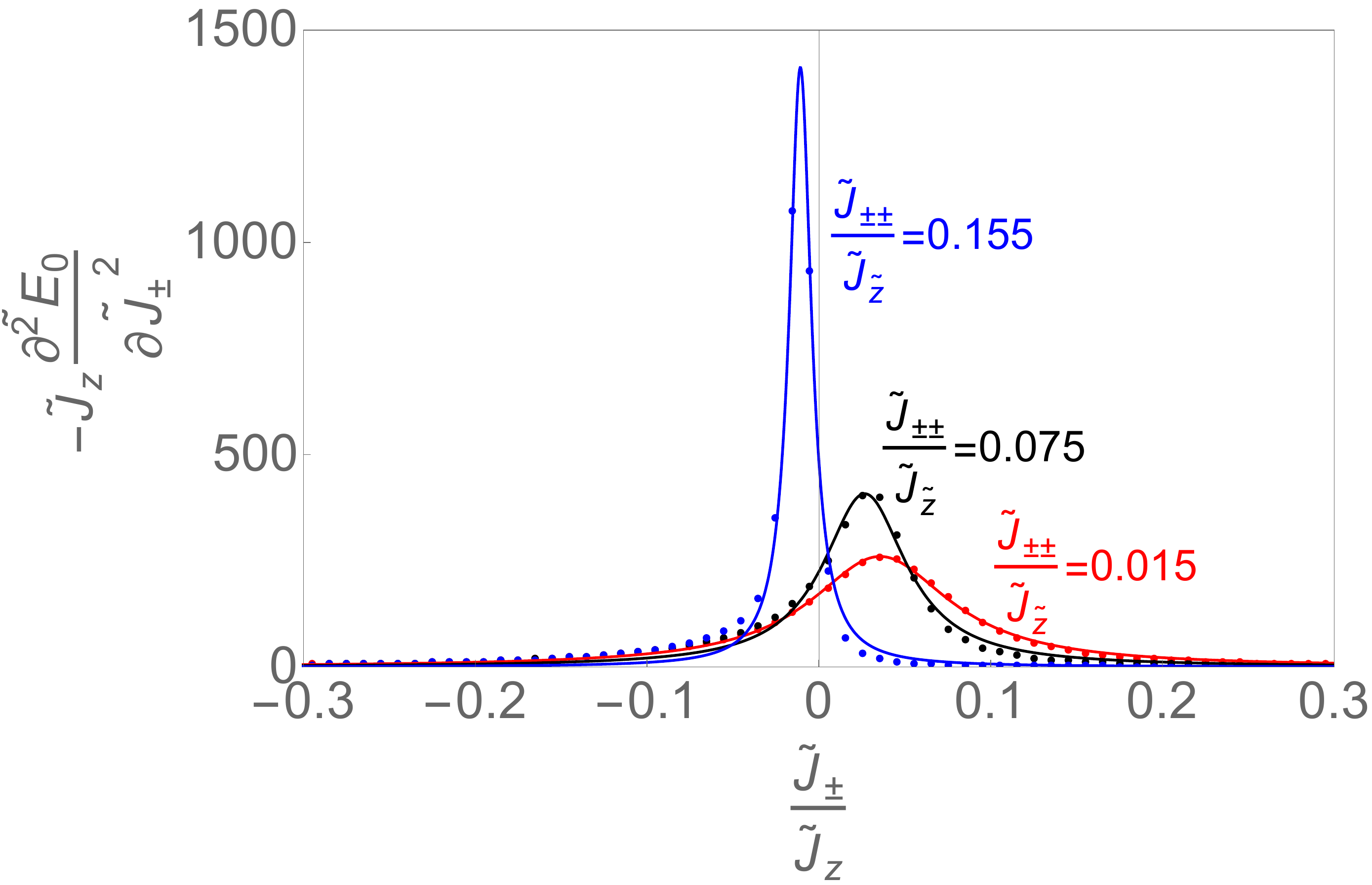}} 
\quad\qquad
\subfigure[\qquad\qquad\qquad]{\includegraphics[width=0.3\textwidth]{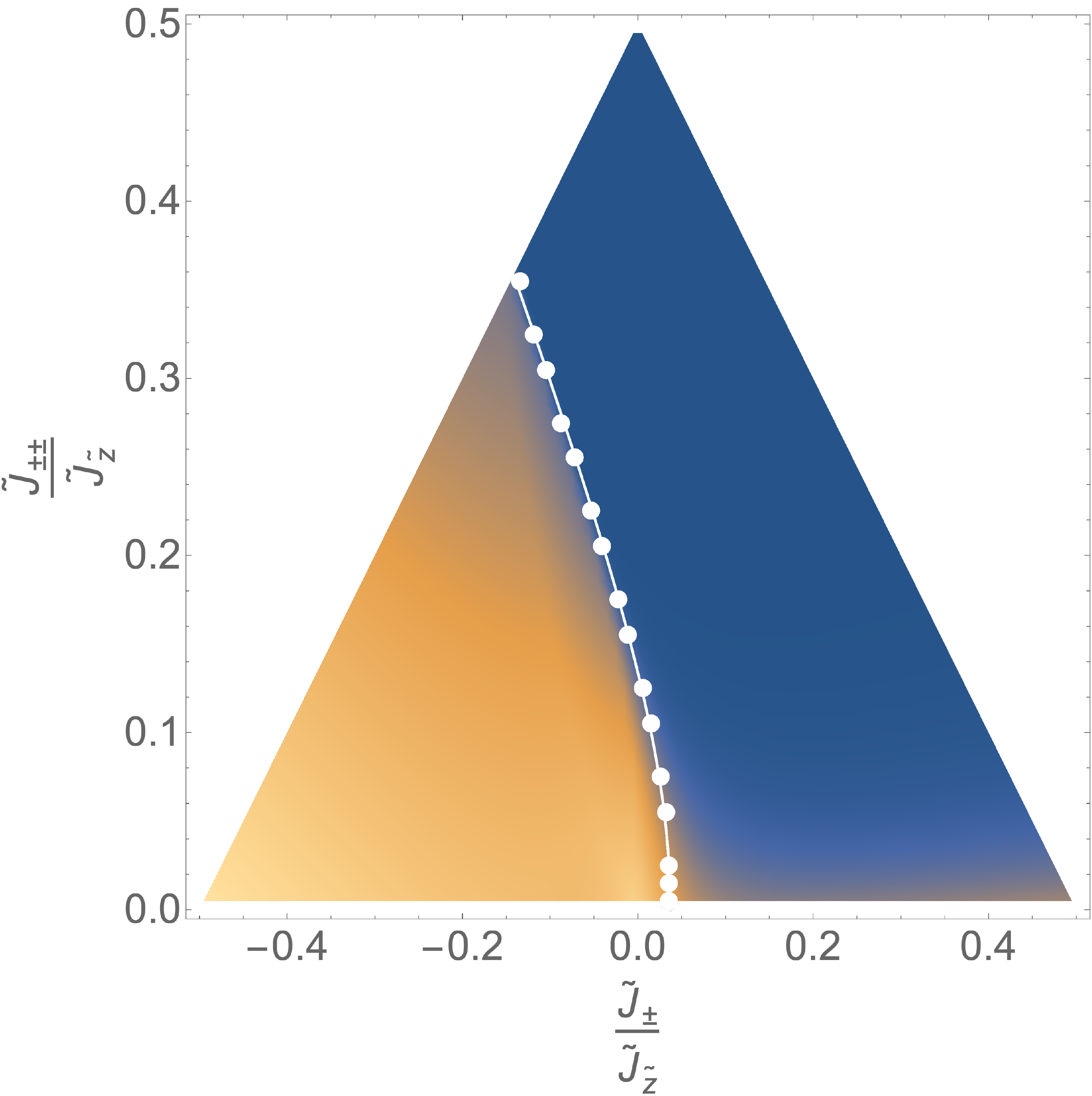} 
\includegraphics[width=0.11\textwidth]{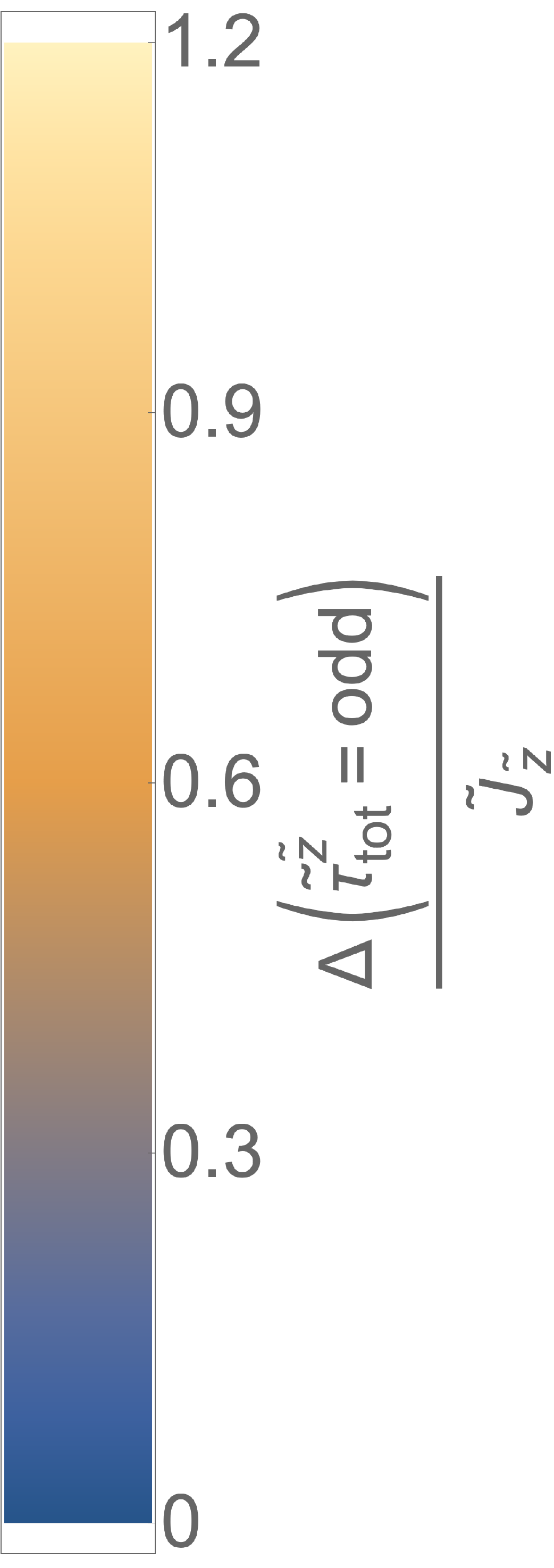}}
\caption{ED calculations of the ground state phase diagram of $\mathcal{H}$
[Eqs. (\ref{eq:HXYZ}), (\ref{eq:Hpm})] within 
the region of parameter space given by (\ref{eq:subregionpm}) with $\tilde{J}_{\tilde{z}}>0$.
(a) Second derivative of the ground state energy in ED on a 16-site cubic cluster with respect to $\tilde{J}_{\pm}$
for various values of $\tilde{J}_{\pm\pm}/\tilde{J}_{\tilde{z}}$.
The peaks indicate a qualitative change in the ground state \cite{chaloupka10}, associated to the transition to long range order.
(b) Color plot of the gap to excitations with odd total $\tilde{\tau}^{\tilde{z}}$ within 16-site ED. The white line 
indicates the
position of peaks in the second derivative of ground state energy [(a)].
The gap collapses rapidly upon crossing the white line, supporting the conclusion that this line corresponds
to a transition to long range order breaking $\pi$-rotation symmetry around the $\tilde{z}$-axis in the thermodynamic limit.
}
\label{fig:ED}
\end{figure*}

\subsection{Cluster variational (CVAR) calculation}
\label{sec:CVAR}

To resolve the CMFT degeneracy in the disordered regime, we follow the cluster variational (CVAR) method 
\cite{benton18-PRL121}. The calculation is described briefly here with further details given in Appendix \ref{sec:CVARapp}.

Labelling CMFT ground states according to their configuration of signs $\{ \sigma \}$ we write down a
generalized superposition of CMFT solutions
\begin{eqnarray}
|\varphi\rangle=\sum_{\{\sigma\}} a_{\{ \sigma \} } | \psi_{\sf CMFT} (\{ \sigma \} )\rangle
\end{eqnarray}
where $a_{\{ \sigma \} } $ are unknown coefficients. We then seek to optimize the new variational energy
\begin{eqnarray}
E_{var}=\frac{\langle \varphi |\mathcal{H}| \varphi \rangle}{\langle \varphi | \varphi \rangle}.
\label{eq:evar}
\end{eqnarray}
Eq. (\ref{eq:evar}) can be expanded in terms of the overlap between distinct CMFT wavefunctions, in a similar
spirit to the derivation of dimer models from an expansion in the overlap between singlet coverings of a lattice
\cite{rokhsar88}.
This generates an effective Hamiltonian in the space of CMFT solutions, where the leading term is a six-site
ring exchange which flips the values of $\sigma_i$
 on hexagonal plaquettes where $\sigma$ alternates in sign around the plaquette, with matrix
 element ${g}_{eff}$.

 This Hamiltonian has already been studied using Quantum Monte Carlo
 \cite{shannon12, benton12-PRB86}.
%     motivated by the fact it is also the Hamiltonian obtained from straightforward perturbation theory in
%     the transverse exchange interactions.
%     %
%     The CVAR expansion provides us a justification for extending the validity of that effective description
%     beyond the strictly perturbative limit
%     \cite{supplemental}.
%     
It can have two different QSL ground states depending on the
sign of ${g}_{eff}$. Both are U(1) QSLs with gapped, bosonic, charges and gapless photons.
The two ground states are distinguished by the U(1) flux threading elementary plaquettes in the ground state.
This background flux vanishes for  ${g}_{eff}<0$ (U(1)$_0$) but is equal to $\pi$ on every plaquette for ${g}_{eff}>0$
(U(1)$_{\pi}$).
The value of  ${g}_{eff}$ can be extracted from the CMFT calculation for all values of exchange parameters (see Appendix
\ref{sec:CVARapp}), and by
this means the degenerate region within CMFT can be divided into two ground state QSL phases (U(1)$_0$ and U(1)$_{\pi}$) depending on the sign of ${g}_{eff}$.
The boundary between regions with different signs of $g_{eff}$ is shown in Fig. \ref{fig:numerics}.
This constitutes our estimate of the boundary between 0-flux and $\pi$-flux QSLs.

\section{Exact Diagonalization}
\label{sec:ED}

\begin{figure*}
\centering
\subfigure[]{\includegraphics[width=0.4\textwidth]{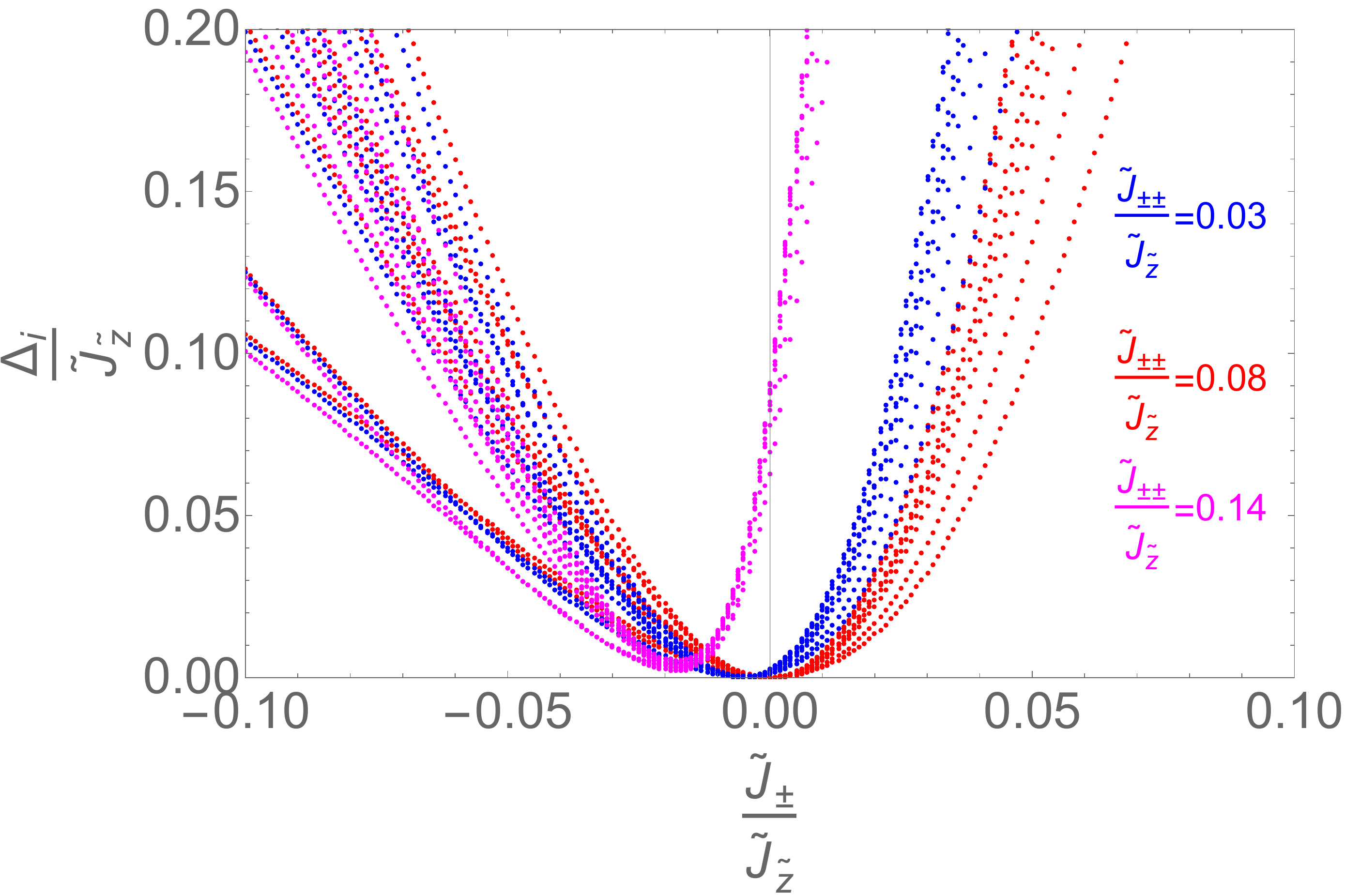}}
\subfigure[]{\includegraphics[width=0.4\textwidth]{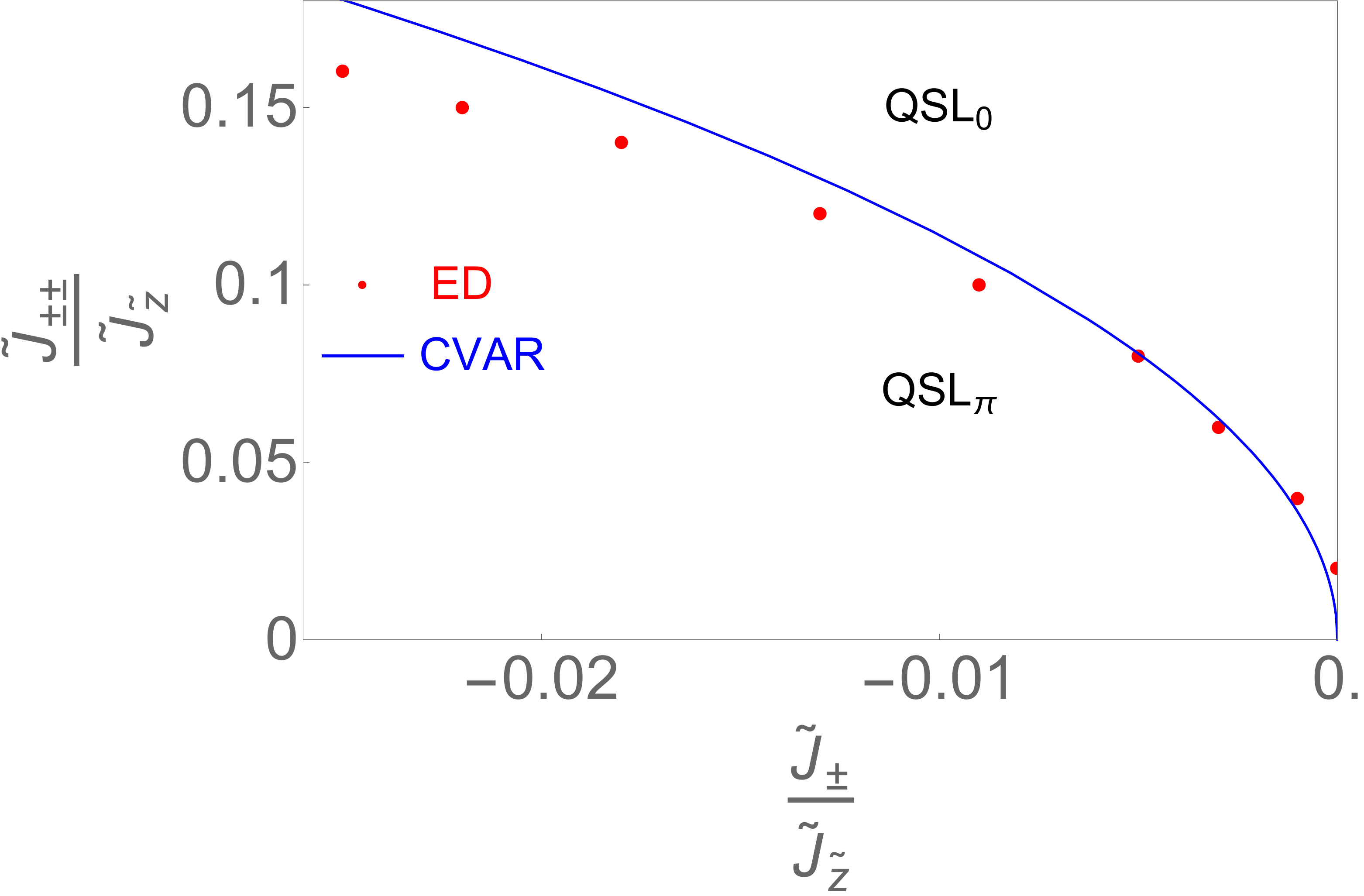}}
\caption{
Estimate of the phase boundary between the quantum spin liquids  QSL$_0$ and QSL$_{\pi}$, within ED.
(a) Collapse of gaps to lowest excited states in the sector with even total $\tilde{\tau}^{\tilde{z}}$, within
16-site ED. The gaps are plotted as a function of  $\tilde{J}_{\pm}/\tilde{J}_{\tilde{z}}$ for three different
values of $\tilde{J}_{\pm\pm}/\tilde{J}_{\tilde{z}}$. For a given value of $\tilde{J}_{\pm\pm}/\tilde{J}_{\tilde{z}}$
the gaps come close to zero around the same point, which we take as an indication of the suppression of tunnelling within the low energy manifold of states. This corresponds with expectations from perturbation theory and CVAR 
calculations for the transition between QSL$_0$ and QSL$_{\pi}$, which happens when the leading tunnelling
term between ice-like states changes sign. 
(b) The position of the collective minima in the gaps in the even sector as a function of exchange parameters, which serves as the ED estimate of the boundary between QSLs (points). This is compared with the estimate of the
same phase boundary from CVAR (solid line).}
\label{fig:QSLboundary}
\end{figure*}

We now turn to ED calculations
on a 16-site cubic cluster with periodic boundaries, to obtain alternative estimates
of the phase boundaries.

\subsection{Boundary of octupolar ordered phase}

Fig. \ref{fig:ED}(a) shows 
the second derivative of the ground state energy on this cluster with respect 
to $\tilde{J}_{\pm}$, at various fixed  
values of $\tilde{J}_{\pm\pm}/\tilde{J}_{\tilde{z}}$.
This second derivative exhibits a peak as $\tilde{J}_{\pm}$ is swept, indicating 
a qualitative change in the ground state \cite{chaloupka10}.

Fig. \ref{fig:ED}(b), shows the position of these peaks as a function of 
$\tilde{J}_{\pm}/\tilde{J}_{\tilde{z}}$ and $\tilde{J}_{\pm\pm}/\tilde{J}_{\tilde{z}}$, laid
over a color plot of the gap to excitations with odd total $\tilde{\tau}^{\tilde{z}}$.
The parity  $p=(-1)^{\sum_i \tilde{\tau}^{\tilde{z}}_i}$ is conserved by $\mathcal{H}$,
with the ground state always having $p=1$.
The line of peaks in the second derivative of the ground state energy coincides with
a rapid decrease of the gap to $p=-1$ excitations. 
This suggests the formation of a twofold degenerate ground state in the thermodynamic limit, 
breaking $\pi$ rotation
symmetry around the $\tilde{z}$ axis, consistent
with the octupolar order identified in CMFT.
We thus interpret the peaks in the second derivative of the ground state energy
as indicative of a transition to octupolar order.

\subsection{Boundary between QSLs}

It is not easy to cleanly distinguish between the two QSL phase, QSL$_0$ and QSL$_{\pi}$ using
ED on a small cluster. However, some insight into how to identify the phase boundary can be gained
by considering how this transition occurs in the perturbative limit and in the CVAR approach.

From the perspective of both perturbation theory and CVAR, the transition from 
QSL$_0$ to QSL$_{\pi}$ occurs when the leading tunnelling matrix element, $g_{eff}$, between ice-like states
changes sign.
At the point where $g_{eff}$ vanishes, tunnelling is restricted to higher order processes and will therefore
be suppressed, leading to a near restoration of the degeneracy of ice-like states.

Returning to ED, this suggests that the transition from QSL$_0$ to QSL$_{\pi}$  will be accompanied by
a simultaneous collapse of many excited states, in the sector with even total $\tilde{\tau}^{\tilde{z}}$, 
to near zero energy.
Such a collapse is indeed observed in the ED data, as shown in Fig. \ref{fig:QSLboundary}(a).
The position of this collective minimum in the gaps within the even sector constitutes the ED estimate
of the phase boundary between the two QSLs.
The phase boundary thus obtained is compared with that from CVAR in Fig. \ref{fig:QSLboundary}(b),
with the two estimates agreeing closely.

\subsection{Combining information from CMFT/CVAR and ED}

Combining the information from  CMFT/CVAR  and ED gives the phase
diagram shown in Fig. \ref{fig:final-pd}. 

For $\tilde{J}_{\pm}<0$ the CMFT and ED estimates of the octupolar
phase boundary agree closely.
For $\tilde{J}_{\pm}>0$ the ED estimates a larger region of octupolar
order (and hence a smaller QSL region) than does the CMFT approach.

For $\tilde{J}_{\pm}>0$ the model has no sign problem from the 
perspective of QMC, and in this regime we can compare with
previous QMC studies.
Several previous QMC studies of the case $\tilde{J}_{\pm\pm}=0$ 
have observed the transition from QSL$_0$ to the ordered phase as 
$\tilde{J}_{\pm}$ is increased \cite{banerjee08, kato15, huang18, huang-arxiv}.
A recent QMC study by Huang {\it et al} \cite{huang-arxiv} has studied the behavior
of this phase boundary as a function of $\tilde{J}_{\pm\pm}$.
Comparison with these results can be used to adjudicate between ED and CVAR where
they disagree.
The ED calculation gives closer agreement with the QMC results from [\onlinecite{huang-arxiv}]
than  CMFT/CVAR does, and therefore we will take the ED calculation as our estimate
of the boundary of the octupolar phase.

The estimates of the boundary between the two QSL phases agree closely between
CVAR  and ED, as shown in Fig. \ref{fig:QSLboundary} (b).
There is, however, some difference between the two estimates at larger negative values of
 $\tilde{J}_{\pm}$. For the purpose of Fig.  \ref{fig:final-pd} we use the boundary from CVAR
 because it gives a more direct prediction of the transition between the two states in the
 thermodynamic limit, as opposed to the more indirect inference from the behavior of gaps in
 ED.

\section{Construction of complete phase diagram}
\label{sec:fullpd}

\begin{figure}
\centering
\includegraphics[width=\columnwidth]{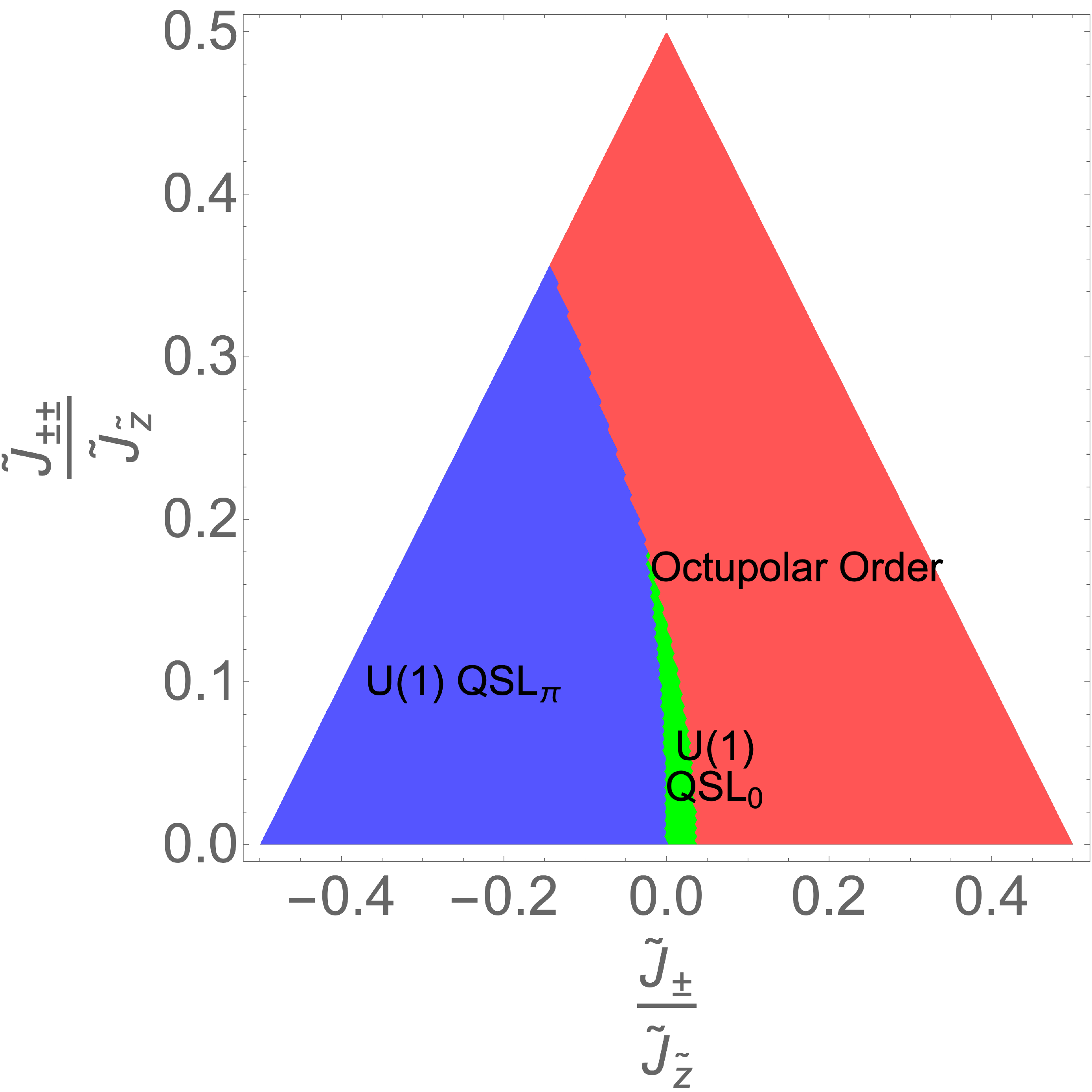}
\caption{Ground state phase diagram of $\mathcal{H}$ [Eqs. (\ref{eq:HXYZ}), (\ref{eq:Hpm})], in the
parameter region (\ref{eq:subregionpm}) with $\tilde{J}_{\tilde{z}}>0$.,
obtained from combining CMFT, CVAR and ED calculations.
This region of parameter space corresponds to the lower region bounded by dashed lines in
Fig. \ref{fig:pd-full}. 
The full phase diagram in Fig. \ref{fig:pd-full} can be generated by applying the dualities
described in Sec \ref{sec:dualities} to this region and to the upper region bounded by dashed lines
in Fig.  \ref{fig:pd-full}  which has all-in-all-out order throughout.}
\label{fig:final-pd}
\end{figure}

The phase diagram in  Fig. \ref{fig:final-pd} can then be extended to the full
parameter space using the duality relations [Eq. (\ref{eq:dualities})].

In doing this, we must take into account how the duality transformations
act on the ground states.
For example, the octupolar ordered phase with $\langle \tilde{\tau}^{\tilde{y}}_i \rangle \neq 0$
becomes an AIAO phase when acted on by a transformation which swaps the
$\tilde{y}$-axis with the $\tilde{x}$-axis or $\tilde{z}$-axis.
On the other hand, transformations which swap only the $\tilde{x}$-axis and $\tilde{z}$-axis,
don't change the classification of the ground state phase because $\tilde{\tau}^{\tilde{x}}_i $
and $\tilde{\tau}^{\tilde{z}}_i $ transform equivalently under point-group and time reversal
symmetries.

Similar considerations allow us to distinguish four different kinds of U(1) QSL, generated from
the two in the phase diagram of Fig. \ref{fig:final-pd}.
In the 0-flux and $\pi$-flux QSLs in Fig. \ref{fig:final-pd} the emergent electric field
of the QSL $E_i \sim \tilde{\tau}^{\tilde{z}}_i$ \cite{hermele04PRB, savary12} and therefore transforms like a magnetic dipole.
If we act a transformation that swaps the $\tilde{z}$-axis and $\tilde{y}$-axes,
then $E_i \sim \tilde{\tau}^{\tilde{y}}_i$ and  transforms like an octupole.
We should therefore not only distinguish U(1) QSLs by the flux but by the dipolar or
octupolar character of the emergent electric field, giving four distinct QSLs on the
complete phase diagram \cite{li17, li19}.

Applying these arguments to the parameter space covered in Fig. \ref{fig:final-pd}, and to the 
case of ${\tilde{J}}_{\tilde{z}}<0$, allows
us to generate the full phase diagram, shown using spherical coordinates [Eq. (\ref{eq:angular_rep})]
in Fig. \ref{fig:pd-full}.

\section{Summary and Outlook}
\label{sec:summary}
  
 We have thus established a phase diagram for the generic, symmetry allowed, nearest
neighbour exchange Hamiltonian describing dipolar-octupolar (DO) pyrochlores R$_2$M$_2$O$_7$
(R=Ce, Sm, Nd).
The picture we arrive at is an encouraging one for the realization of QSL states.
There are four distinct U(1) QSLs on the phase diagram of the generic nearest neighbor model,
and between them they occupy $\sim19$\% of the available parameter space.
%
%With at least ten DO pyrochlores having already been synthesized (Ce$_2$Zr$_2$O$_7$,
%Ce$_2$Sn$_2$O$_7$, Sm$_2$Zr$_2$O$_7$, Sm$_2$Sn$_2$O$_7$, Sm$_2$Ti$_2$O$_7$, 
%Nd$_2$Zr$_2$O$_7$, Nd$_2$Sn$_2$O$_7$, Nd$_2$Hf$_2$O$_7$, Nd$_2$Pb$_2$O$_7$),
%the chances of realizing a U(1) QSL in at least one of them seem high.

Amongst materials, Ce$_2$Zr$_2$O$_7$ \cite{gao19, gaudet19},
Ce$_2$Sn$_2$O$_7$ \cite{sibille15, sibille-arXiv} and Sm$_2$Zr$_2$O$_7$ \cite{xu-thesis}
stand out as lacking low temperature order.
The Ce pyrochlores in particular seem promising with recent neutron scattering results
on Ce$_2$Zr$_2$O$_7$ bearing similarity to predictions for emergent photons \cite{gaudet19}.
Low energy correlations in 
Ce$_2$Sn$_2$O$_7$ seem to be dominantly octupolar in nature \cite{sibille-arXiv}, which would
be consistent with either of the two octupolar spin liquids on the phase diagram 
[Fig. \ref{fig:pd-full}].

It will be important to establish estimates of the exchange parameters of 
Ce$_2$Zr$_2$O$_7$ and Ce$_2$Sn$_2$O$_7$, combining information from inelastic
neutron scattering with fits to thermodynamic data.
Mean field calculations in [\onlinecite{sibille-arXiv}] give an initial estimate for Ce$_2$Sn$_2$O$_7$
of $J_y=0.48 {\rm K}, J_z=0.03 {\rm K}$, while setting $J_x$ and $J_{xz}$  to zero, in the basis of Eq.  (\ref{eq:H0}).
This would place Ce$_2$Sn$_2$O$_7$ in the Octupolar-$U(1)_{\pi}$ region of the
phase diagram.
It would be useful to refine this estimate with all parameters allowed
to be finite, and using calculations beyond mean field theory.

If refined parameterisations place Ce$_2$Zr$_2$O$_7$ and Ce$_2$Sn$_2$O$_7$ within
the QSL regimes of Fig. \ref{fig:pd-full}, then this will be a strong indication that they are indeed
U(1) QSLs, and the parameterized model will provide a platform for further theoretical study.
Understanding the effects of disorder of the crystal structure is also likely to be crucial, particularly
in regard to the possible substitution of magnetic Ce$^{3+}$ with non-magnetic Ce$^{4+}$ \cite{gaudet19}.

For those DO pyrochlores that are known to possess magnetic order at low temperature, the spin
liquid phases may also manifest at finite temperature, as suggested recently in Nd$_2$Zr$_2$O$_7$ \cite{xu-arxiv}.
In such cases it may even be possible to tune into the $T=0$ QSL phase using chemical
or physical pressure, giving another avenue to realize these exotic states of matter.\\

{\it Note:}
After completion of this work, the author became aware of a recent paper by Patri {\it et al}  \cite{patri-arxiv} which also
presents calculations of the ground state phase diagram of DO pyrochlores.\\

{\it Acknowledgements:}
The author acknowledges useful discussions with Andrea Bianchi, 
Jonathan Gaudet, Bruce Gaulin, Ludovic Jaubert,
Bella Lake, Kate Ross, Nic Shannon, Romain Sibille, Rajiv Singh, Evan Smith,  Jianhui Xu
and Danielle Yahne.
The author also thanks
Paul McClarty for comments on the draft manuscript.

\appendix

\section{CMFT solutions in the ice-like r{\'e}gime}
\label{sec:CMFTapp}

\begin{figure}
\centering
\includegraphics[width=0.3\textwidth]{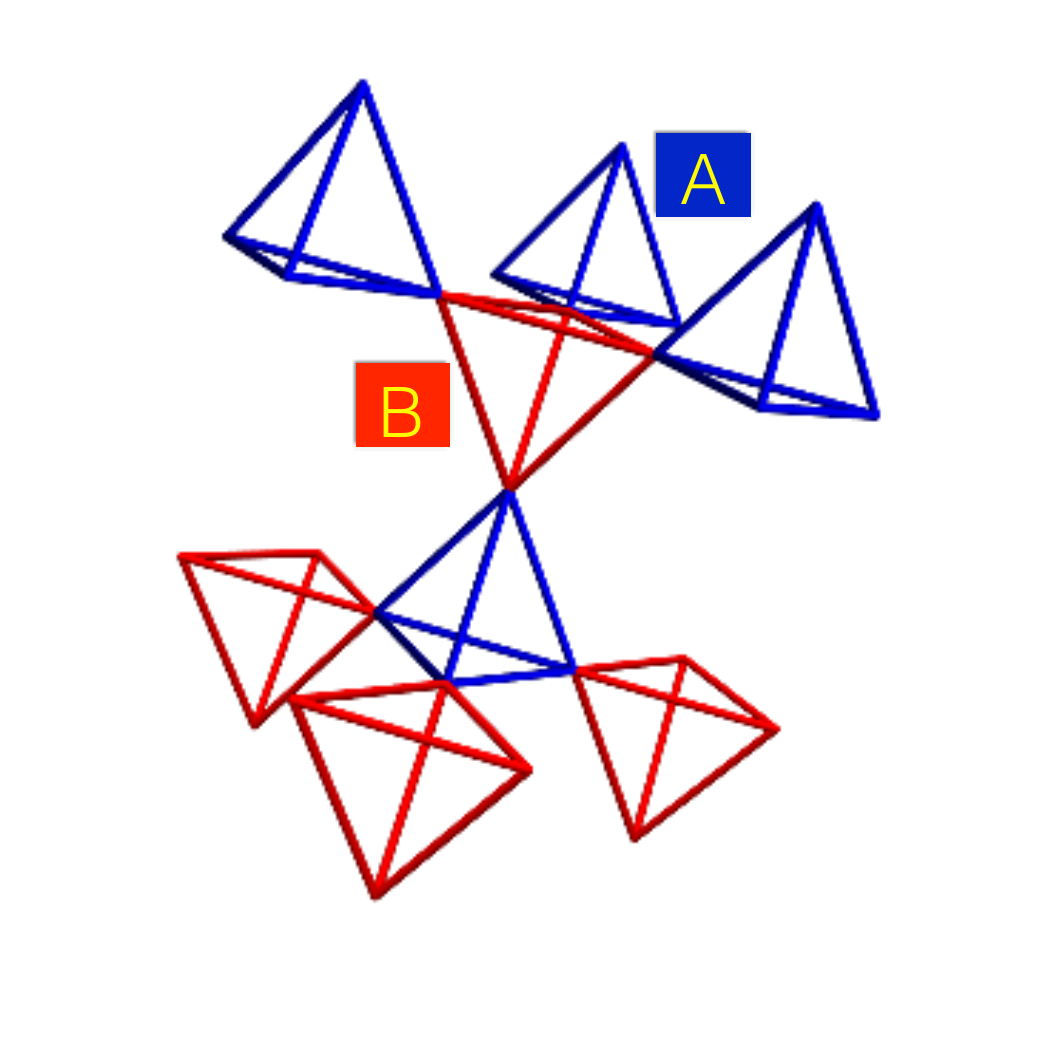}
\caption{The pyrochlore lattice with `A' and `B' tetrahedra highlighted in red and blue, respectively.}
\label{fig:ABtet}
\end{figure}

The CMFT proceeds by optimizing variational wavefunctions of the form:
\begin{eqnarray}
|\psi_{\sf CMFT} ( \{ {\bf h}_i \} )\rangle=\prod_{t \in A} |\phi_t ( \{ {\bf h}_{i \in t} \} ) \rangle
\label{eq:psi_cmft}
\end{eqnarray}
where the product is over all `A' tetrahedra [Fig. \ref{fig:ABtet}], $ \{ {\bf h}_i \}$ is the configuration of auxiliary fields
defined on each site and the single tetrahedron wavefunctions $|\phi_t\rangle$ depend only on the fields
on sites belonging to tetrahedron $t$.
The auxiliary fields $ {\bf h}_i$ are variational parameters for optimizing the CMFT energy
\begin{eqnarray}
E_{\sf CMFT}=
\langle \psi_{\sf CMFT} |
\mathcal{H}
|
\psi_{\sf CMFT} \rangle
\label{eq:ECMFT}
\end{eqnarray}

The wave functions $ |\phi_t ( \{ {\bf h}_{i \in t} \} )$ are taken to be eigenstates of a single tetrahedron
Hamiltonian $\mathcal{H}'_t$
\begin{eqnarray}
&&\mathcal{H}'_t=\sum_{\langle ij \rangle \in t} \sum_{\alpha=\tilde{x},{y},{z}} \tilde{J}_{\tilde{\alpha}} 
\tilde{\tau}_i^{\tilde{\alpha}} \tilde{\tau}_j^{\tilde{\alpha}}-
\sum_{i \in t} \sum_{\alpha=\tilde{x},{y},{z}} h_i^{\alpha} \tilde{\tau}_i^{{\alpha}}
\label{eq:Hprime}\\
&&\mathcal{H}'_t |\phi_t ( \{ {\bf h}_{i \in t} \} ) \rangle=\epsilon_{0,t} |\phi_t ( \{ {\bf h}_{i \in t} \} ) \rangle.
\label{eq:MFTeigval}
\end{eqnarray}

$E_{\sf CMFT}$ is then:
\begin{eqnarray}
&&E_{\sf CMFT}=\sum_{t \in A} \epsilon_{0,t}+
\sum_{i} \sum_{\alpha=\tilde{x},{y},{z}} h_i^{\alpha} \langle \tilde{\tau}_i^{{\alpha}} \rangle \nonumber \\
&&\qquad
+ \sum_{\langle ij \rangle_B} \sum_{\alpha=\tilde{x},{y},{z}} \tilde{J}_{\tilde{\alpha}} 
\langle \tilde{\tau}_i^{\tilde{\alpha}} \rangle \langle \tilde{\tau}_j^{\tilde{\alpha}} \rangle
\label{eq:ECMFT-2}
\end{eqnarray}
where the final term in Eq. (\ref{eq:ECMFT-2}) sums over bonds belonging to `B' tetrahedra
and accounts for the interactions on those tetrahedra.

There is a large region of the phase diagram [Fig. 2 of main text] in which the optimal
solutions for ${\bf h}_{i}$ take the form:
\begin{eqnarray}
&&h_i^{\tilde{x}}=h_i^{\tilde{y}}=0,  h_z^{\tilde{z}}=\sigma_i h \nonumber \\
&&\sigma_i=\pm1.
\label{eq:h-spinice}
\end{eqnarray}
Correspondingly, the expectation values of the spin components are:
\begin{eqnarray}
&&\langle \tilde{\tau}_i^{\tilde{x}} \rangle = \langle \tilde{\tau}_i^{\tilde{y}} \rangle=0 \nonumber \\
&&\langle \tilde{\tau}_i^{\tilde{z}} \rangle =\sigma_i s
\end{eqnarray}
with $h$ and $s$ being uniform across the system, and fixed by the energy optimization for a given
parameter set.

With this form for the auxiliary fields, the mean field energy [Eq. (\ref{eq:ECMFT-2})] becomes
\begin{eqnarray}
E_{\sf CMFT}=\sum_{t \in A} \epsilon_{0,t}+
N h s
+ \tilde{J}_{\tilde{z}} s^2
\sum_{\langle ij \rangle_B} \sigma_i \sigma_j.
\label{eq:ECMFT-SI}
\end{eqnarray}
Any arrangement of signs $\sigma_i$ such that
\begin{eqnarray}
\sum_{i \in t} \sigma_i=0 \quad \forall \ {\rm tetrahedra} \ t
\label{eq:icerule}
\end{eqnarray}
gives rise to the same value of $\epsilon_{0,t}$, as can be inferred from the symmetries of the original Hamiltonian. The remaining terms in Eq. (\ref{eq:ECMFT-SI}) are also the 
same for all configurations obeying Eq. (\ref{eq:icerule}).
Thus we have a large degeneracy of mean field solutions in this regime.

\begin{figure*}
\centering
\includegraphics[width=0.8\textwidth]{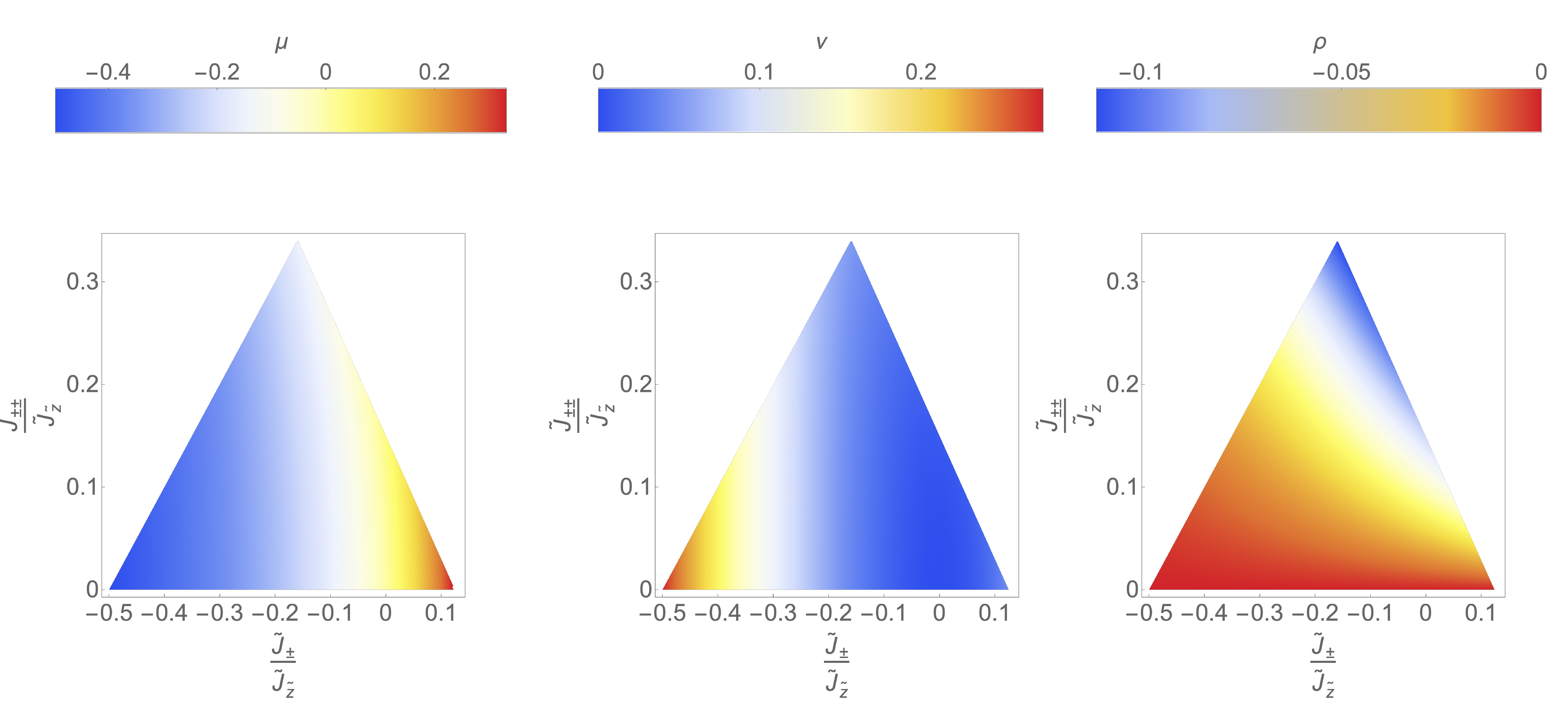}
\caption{Parameters $\mu$, $\nu$ and $\rho$ which enter the single tetrahedron wavefunctions
[Eq. (\ref{eq:wf-explicit})], plotted as a function of the exchange parameters in the region where the
CMFT solutions have an ice-like degeneracy.}
\label{fig:wf-params}
\end{figure*}

Each arrangement of signs $\sigma_i$ obeying Eq. (\ref{eq:icerule}) defines a CMFT wavefunction [via 
Eqs.  (\ref{eq:psi_cmft}), (\ref{eq:MFTeigval}) and (\ref{eq:h-spinice})]
which we will denote with $|\psi_{CMFT} ( \{\sigma\}) \rangle $.
Explicitly, the form of single tetrahedron wave functions 
$| \phi_t (\{\sigma_{i\in t})\rangle$ (denoted simply as $| \sigma_0 \sigma_1 \sigma_2 \sigma_3 \rangle$)
relates to the configuration of signs 
on $t$ in the following way, written in the basis diagonalizing $\tilde{\tau}_i^{\tilde{z}}$:
\begin{eqnarray}
&& | + + - -  \rangle = \sqrt{1-\mu^2-\nu^2-\rho^2} | \uparrow \uparrow \downarrow \downarrow  \rangle
\nonumber \\
&& 
\qquad + \frac{\mu}{2} 
\left(
| \uparrow  \downarrow  \uparrow \downarrow  \rangle +
| \uparrow  \downarrow \downarrow   \uparrow \rangle  +
| \downarrow \uparrow \uparrow  \downarrow  \rangle +
|\downarrow \uparrow   \downarrow   \uparrow \rangle
\right) 
\nonumber \\
&& 
\qquad 
+\nu
|\downarrow    \downarrow \uparrow  \uparrow \rangle
+ \frac{\rho}{\sqrt{2}}
\left(
|\uparrow \uparrow   \uparrow   \uparrow \rangle
+|\downarrow    \downarrow  \downarrow \downarrow \rangle
\right) \nonumber\\
&& | + - + -  \rangle = \sqrt{1-\mu^2-\nu^2-\rho^2} | \uparrow  \downarrow \uparrow \downarrow  \rangle
\nonumber \\
&& 
\qquad 
+ \frac{\mu}{2} 
\left(
| \uparrow    \uparrow \downarrow \downarrow  \rangle +
| \uparrow  \downarrow \downarrow   \uparrow \rangle  +
| \downarrow \uparrow \uparrow  \downarrow  \rangle +
|\downarrow    \downarrow  \uparrow  \uparrow \rangle
\right) 
\nonumber \\
&& 
\qquad 
+\nu
|\downarrow \uparrow   \downarrow   \uparrow \rangle
+ \frac{\rho}{\sqrt{2}}
\left(
|\uparrow \uparrow   \uparrow   \uparrow \rangle
+|\downarrow    \downarrow  \downarrow \downarrow \rangle
\right) \nonumber\\
&& | + -  - +  \rangle = \sqrt{1-\mu^2-\nu^2-\rho^2} | \uparrow  \downarrow  \downarrow  \uparrow \rangle
\nonumber \\
&& 
\qquad + \frac{\mu}{2} 
\left(
| \uparrow    \uparrow \downarrow \downarrow  \rangle +
| \uparrow  \downarrow    \uparrow \downarrow \rangle  +
| \downarrow \uparrow   \downarrow \uparrow \rangle +
|\downarrow    \downarrow  \uparrow  \uparrow \rangle
\right) 
\nonumber \\
&& 
\qquad 
+\nu
|\downarrow \uparrow      \uparrow  \downarrow \rangle
+ \frac{\rho}{\sqrt{2}}
\left(
|\uparrow \uparrow   \uparrow   \uparrow \rangle
+|\downarrow    \downarrow  \downarrow \downarrow \rangle
\right) \nonumber\\
&& |  - - ++  \rangle = \sqrt{1-\mu^2-\nu^2-\rho^2} |  \downarrow \downarrow \uparrow \uparrow  \rangle
\nonumber \\
&& 
\qquad + \frac{\mu}{2} 
\left(
| \uparrow  \downarrow  \uparrow \downarrow  \rangle +
| \uparrow  \downarrow \downarrow   \uparrow \rangle  +
| \downarrow \uparrow \uparrow  \downarrow  \rangle +
|\downarrow \uparrow   \downarrow   \uparrow \rangle
\right) 
\nonumber \\
&& 
\qquad 
+\nu
| \uparrow  \uparrow \downarrow    \downarrow \rangle
+ \frac{\rho}{\sqrt{2}}
\left(
|\uparrow \uparrow   \uparrow   \uparrow \rangle
+|\downarrow    \downarrow  \downarrow \downarrow \rangle
\right) \nonumber
\end{eqnarray}

\begin{eqnarray}
&& |-+-+  \rangle = \sqrt{1-\mu^2-\nu^2-\rho^2} | \downarrow  \uparrow \downarrow \uparrow  \rangle
\nonumber \\
&& 
\qquad + \frac{\mu}{2} 
\left(
| \uparrow    \uparrow \downarrow \downarrow  \rangle +
| \uparrow  \downarrow \downarrow   \uparrow \rangle  +
| \downarrow \uparrow \uparrow  \downarrow  \rangle +
|\downarrow    \downarrow  \uparrow  \uparrow \rangle
\right) \nonumber \\
&& 
\qquad 
+\nu
| \uparrow \downarrow     \uparrow \downarrow \rangle
+ \frac{\rho}{\sqrt{2}}
\left(
|\uparrow \uparrow   \uparrow   \uparrow \rangle
+|\downarrow    \downarrow  \downarrow \downarrow \rangle
\right) \nonumber\\
&& | -++- \rangle = \sqrt{1-\mu^2-\nu^2-\rho^2} |   \downarrow \uparrow\uparrow \downarrow   \rangle
\nonumber \\
&& 
\qquad + \frac{\mu}{2} 
\left(
| \uparrow    \uparrow \downarrow \downarrow  \rangle +
| \uparrow  \downarrow    \uparrow \downarrow \rangle  +
| \downarrow \uparrow   \downarrow \uparrow \rangle +
|\downarrow    \downarrow  \uparrow  \uparrow \rangle
\right) 
\nonumber \\
&& 
\qquad 
+\nu
|\uparrow\downarrow         \downarrow\uparrow \rangle
+ \frac{\rho}{\sqrt{2}}
\left(
|\uparrow \uparrow   \uparrow   \uparrow \rangle
+|\downarrow    \downarrow  \downarrow \downarrow \rangle
\right) 
\label{eq:wf-explicit}
\end{eqnarray}
The parameters $\mu, \nu$ and $\rho$ can always be chosen to be real.
This choice, combined with the choice to define the first term on the right hand side of each
line of (\ref{eq:wf-explicit}) to be positive, removes any phase ambiguity in the CMFT wavefunctions.
$\mu, \nu$ and $\rho$   vary as a function of the
exchange parameters $ \tilde{J}_{\tilde{\alpha}} $
and are plotted in Fig. \ref{fig:wf-params}.

\section{Details of CVAR calculation}
\label{sec:CVARapp}

\begin{figure}
\centering
\includegraphics[width=\columnwidth]{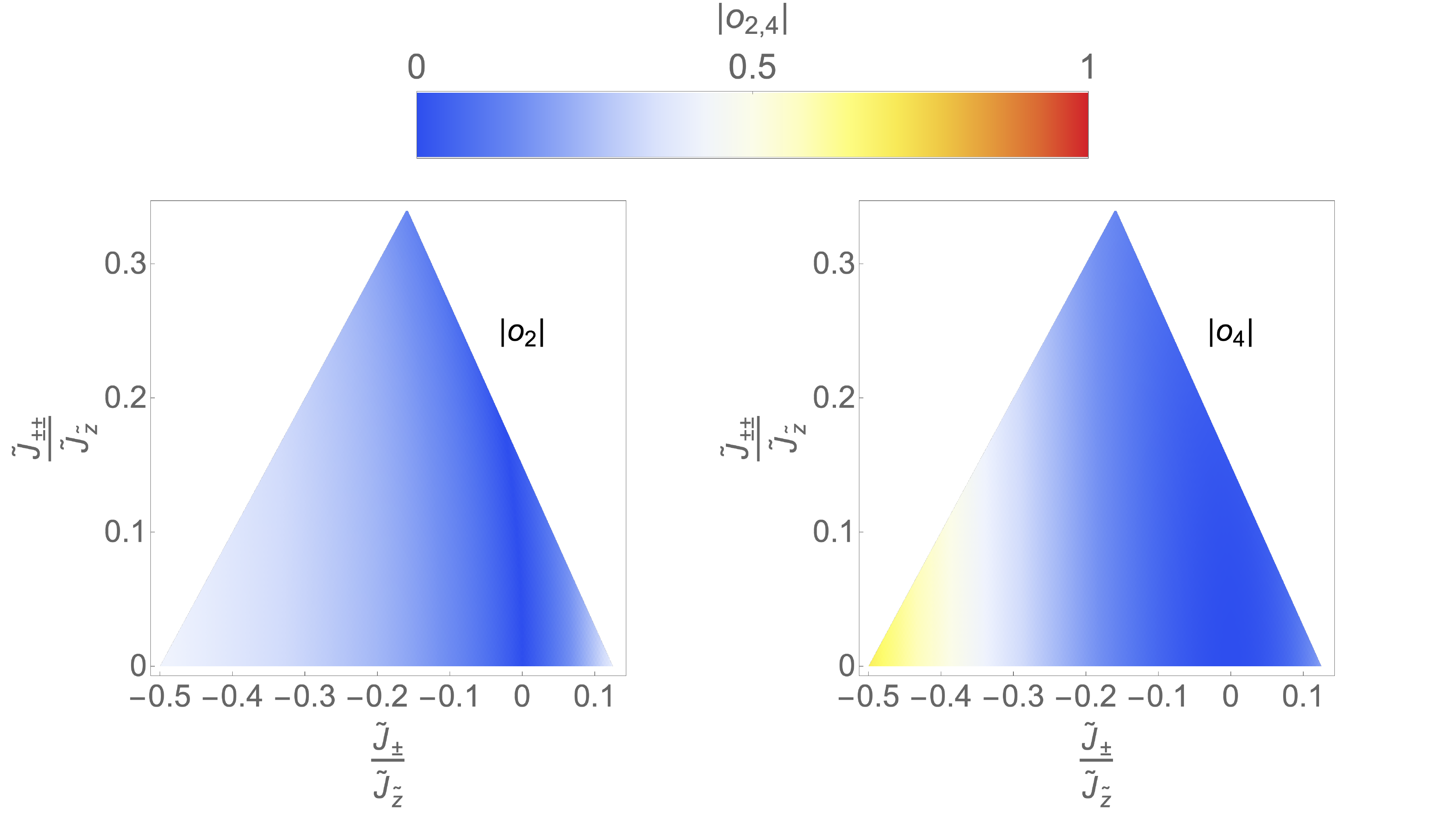}
\caption{Magnitudes of the single tetrahedron overlap parameters $o_2$ and $o_4$.
These function as the small parameters for the expansion of the CVAR energy.}
\label{fig:wf-overlaps}
\end{figure}

The goal of the CVAR calculation is to resolve the degeneracy of the CMFT solutions by 
considering a new trial wavefunction which is a superposition of the CMFT solutions:
\begin{eqnarray}
|\varphi\rangle=\sum_{\{\sigma\}} a_{\{\sigma\}} |\psi_{\sf CMFT} ( \{ \sigma \} ) \rangle.
\label{eq:wf-cvar}
\end{eqnarray}
where $a_{\{\sigma\}}$ are, {\it a priori} unknown, complex, coefficients.

We then seek to optimize the variational energy
\begin{eqnarray}
&&E_{var}=\frac{\langle \varphi | \mathcal{H} | \varphi \rangle}{\langle \varphi | \varphi \rangle}=
\frac{\sum_{\{\sigma\} \{\sigma'\}}  a^{\ast}_{\{\sigma'\}}  a_{\{\sigma\}}  X_{{\{\sigma'\}} {\{\sigma\}}}}
{\sum_{\{\sigma\} \{\sigma'\}}  a^{\ast}_{\{\sigma'\}}  a_{\{\sigma\}}  O_{{\{\sigma'\}} {\{\sigma\}}}}
\nonumber \\
&& \qquad
\equiv
\frac{{\bf a}^{\dagger} \cdot {\bf X} \cdot {\bf a}}{{\bf a}^{\dagger} \cdot {\bf O} \cdot {\bf a}}
\end{eqnarray}
where ${\bf X}$ is a matrix containing the Hamiltonian matrix elements between different
CMFT wavefunctions and ${\bf O}$ contains the overlaps (the CMFT wavefunctions are not generally orthogonal to one another)
\begin{eqnarray}
&&X_{{\{\sigma'\}} {\{\sigma\}}}= \langle \psi_{\sf CMFT} ( \{ \sigma' \} ) | \mathcal{H} |  \psi_{\sf CMFT} ( \{ \sigma' \} ) \rangle\\
&&O_{{\{\sigma'\}} {\{\sigma\}}}=\langle \psi_{\sf CMFT} ( \{ \sigma' \} )|  \psi_{\sf CMFT} ( \{ \sigma' \} ) \rangle.
\end{eqnarray}

It is then useful to define a new matrix ${\bf X}'$ with vanishing diagonal elements:
\begin{eqnarray}
{\bf X}'={\bf X}-E_{\sf CMFT} {\bf O}
\end{eqnarray}
such that
\begin{eqnarray}
E_{var}=E_{\sf CMFT}+\frac{{\bf a}^{\dagger} \cdot {\bf X}' \cdot {\bf a}}{{\bf a}^{\dagger} \cdot {\bf O} \cdot {\bf a}}.
\end{eqnarray}

We then relate the vector of coefficients ${\bf a}$, to a new normalized vector ${\bf b}$ via:
\begin{eqnarray}
&&{\bf a}={\bf O}^{-1/2} \cdot {\bf b} 
\label{eq:ab}
\\
&&{\bf b}^{\dagger} \cdot {\bf b}=1
\end{eqnarray}

The variational energy is then
\begin{eqnarray}
E_{var}=E_{\sf CMFT}+{\bf b}^{\dagger} \cdot {\bf H}_{\sf eff} \cdot {\bf b}
\end{eqnarray}
where
\begin{eqnarray}
{\bf H}_{\sf eff}={\bf O}^{-1/2}  \cdot {\bf X}' \cdot  {\bf O}^{-1/2}
\label{eq:Heff}
\end{eqnarray}
The optimal superposition of CMFT solitions is then given by the ground state of ${\bf H}_{\sf eff}$
and Eq. (\ref{eq:ab}).

\begin{figure}
\centering
\includegraphics[width=\columnwidth]{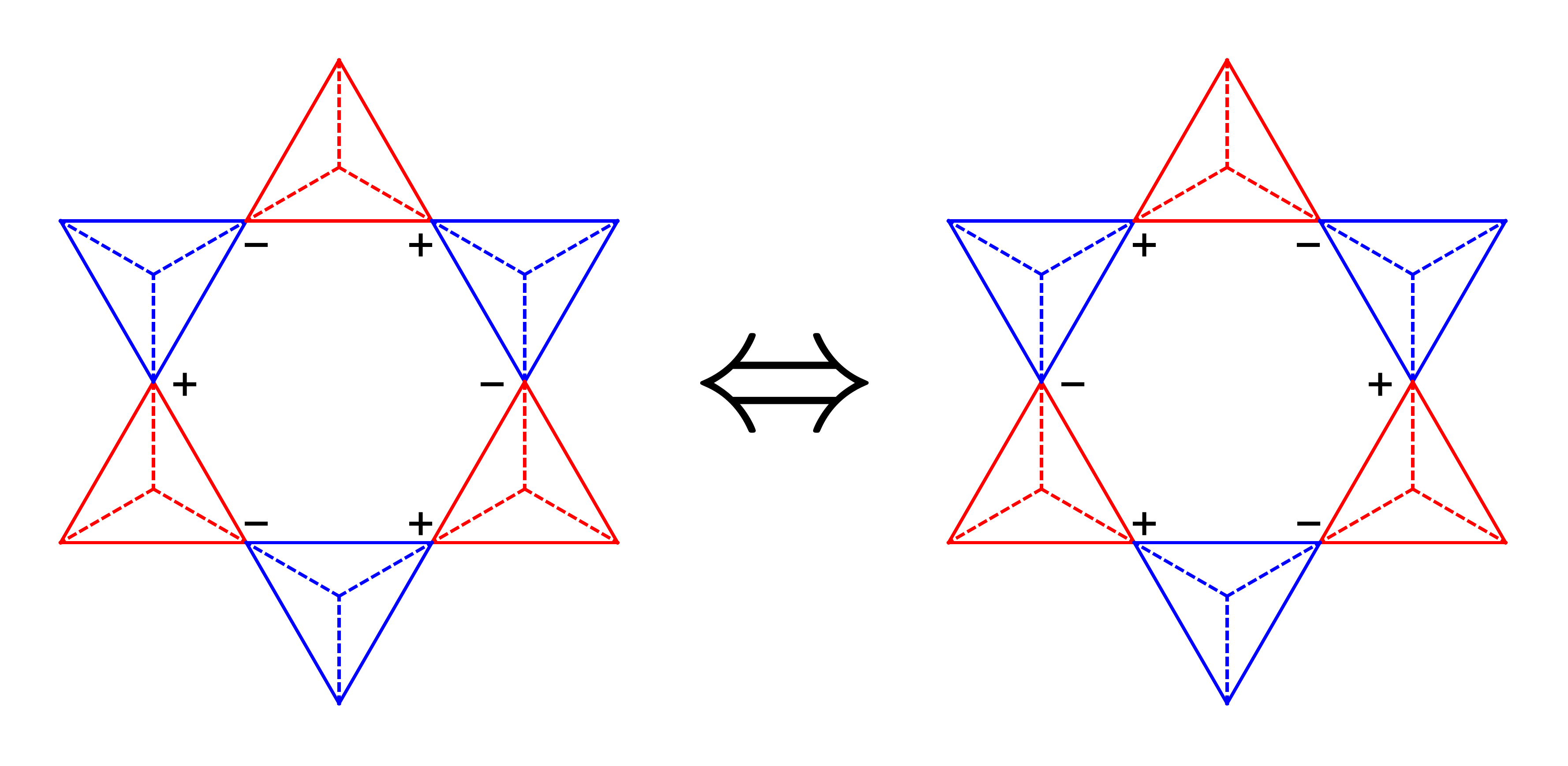}
\caption{
Processes which flip a six-site loop of alternating sign variables $\sigma_i$ provide the
leading matrix elements in ${\bf X}'$,${\bf O}'$ and ${\bf H}_{\sf eff}$ [Eq. (\ref{eq:Heff})], 
\cite{benton18-PRL121}.
}
\label{fig:hexflip}
\end{figure}

We then expand ${\bf H}_{\sf eff}$ in terms of two overlap parameters $o_2$ and $o_4$, 
which can be defined from the wavefunctions in Eqs. (\ref{eq:wf-explicit}):
\begin{eqnarray}
&&o_2=\langle ++--|+-+-\rangle=\frac{\mu^2}{2}
\nonumber \\
&& \qquad
+\rho^2+\mu(\nu+\sqrt{1-\mu^2-\nu^2-\rho^2}) \\
&&o_4=\langle ++--|+-+-\rangle=\mu^2+\nonumber \\
&& \qquad
\rho^2+2\nu \sqrt{1-\mu^2-\nu^2-\rho^2}.
\end{eqnarray}
These two quantities are treated as small parameters for the purposes of the expansion and indeed
they are small through most of the relevant parameter space, as shown in Fig. \ref{fig:wf-overlaps}.

To expand Eq. (\ref{eq:Heff}) we note that all the diagonal elements of ${\bf O}$ are unity, and the leading off diagonal elements $\sim (o_2)^3$ (coming from the process illustrated in Fig. \ref{fig:hexflip}) so we can
write
\begin{eqnarray}
{\bf O}^{-1/2}=(1+{\bf O}')^{-1/2} \approx 1-\frac{1}{2}{\bf O}'.
\end{eqnarray}

The first two terms of the expansion of ${\bf H}_{\sf eff}$ are then:
\begin{eqnarray}
{\bf H}_{\sf eff}\approx{\bf X}'-\frac{1}{2} ({\bf O}'\cdot{\bf X}'+{\bf X}'\cdot{\bf O}').
\end{eqnarray}
The leading elements in ${\bf X}'$ are $\sim(o_2)^2$ (again, from the process in Fig. \ref{fig:hexflip})  and the leading elements in  ${\bf O}'\cdot{\bf X}'$
are $\sim(o_2)^5$ and so we henceforth drop the second term.

We then need to evaluate the leading matrix elements in ${\bf X}'$ which connect configurations of $\sigma_i$ which differ on a single hexagonal plaquette as shown in Fig. \ref{fig:hexflip}.
The matrix element to a flip a hexagon is $g_{eff}$.
The sign of $g_{eff}$ determines whether the ground state should be a 0 or $\pi$ flux QSL, with
\begin{eqnarray}
&&g_{eff}<0 \implies {\rm U(1) QSL_0} \\
&&g_{eff}>0 \implies {\rm U(1) QSL_{\pi}} 
\end{eqnarray}
as may be inferred from prior quantum Monte Carlo studies of the six-site ring exchange Hamiltonian
\cite{shannon12} and from a unitary transformation which relates the sign-problem free case ($g_{eff}<0$)
to the frustrated case  ($g_{eff}>0$) \cite{hermele04PRB}.

\begin{figure}
\includegraphics[width=0.4\textwidth]{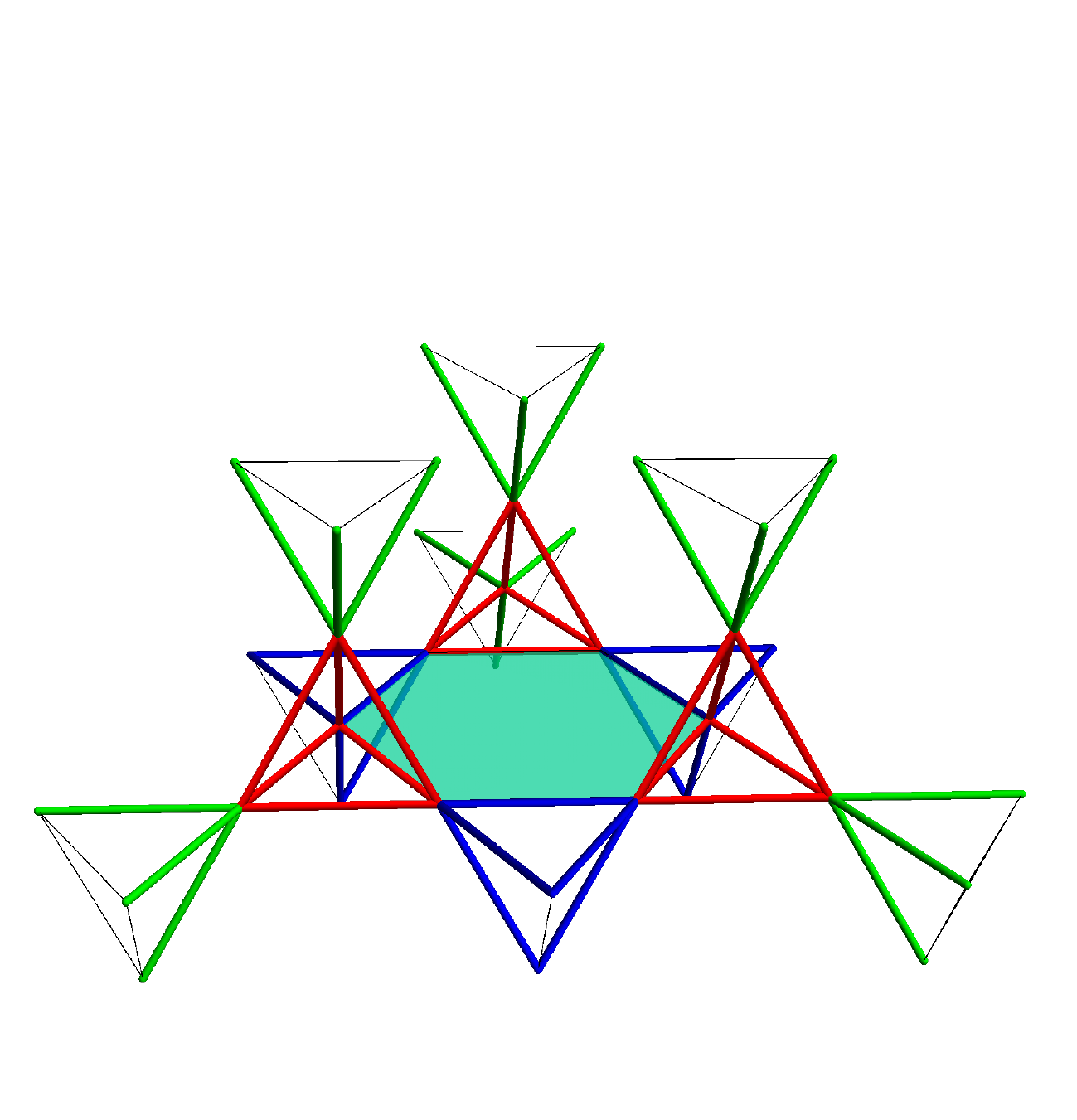}
\caption{Contributions to $g_{eff}$. The matrix element between two configurations which differ by
flipping the central hexagon has three distinct non-vanishing contributions:  from the `A' tetrahedra highlighted in red, from the `B' bonds in blue which connect to the interior of the hexagon and
from the `B' bonds in green which connect to the exterior of  the `A' tetrahedra  belonging to the
hexagon.
The contribution from bonds drawn with narrow black lines vanishes.
}
\label{fig:contributions}
\end{figure}

Quite generally the matrix element of ${\bf X}'$ between two CMFT wavefunctions can be written as
\begin{eqnarray}
&&X'_{\{\sigma'\}\{\sigma\}}= \nonumber \\
&& \qquad
\sum_{t \in A} \left(
\langle \psi_{\sf CMFT} (\{ \sigma'\}) | \mathcal{H}_t   | \psi_{\sf CMFT} (\{ \sigma\}) \rangle   - 
\epsilon_A O_{\{\sigma'\}\{\sigma\}} \right) \nonumber \\
&& \qquad +
\sum_{\langle ij \rangle \in B}  \tilde{J}_{\tilde{z}} \bigg(
 \langle \psi_{\sf CMFT} (\{ \sigma'\}) |  \tilde{\tau}^{\tilde{z}}_i   \tilde{\tau}^{\tilde{z}}_j
 | \psi_{\sf CMFT} (\{ \sigma\}) \rangle   - \nonumber \\
&& \qquad 
 s^2 \sigma_{i} \sigma_j  O_{\{\sigma'\}\{\sigma\}} 
 \bigg)
 \label{eq:matrix-elements}
\end{eqnarray}
where the first sum is over `A' tetrahedra and the second is over bonds belonging to `B' 
tetrahedra.
$\mathcal{H}_t$ is the original exchange Hamiltonian on tetrahedron $t$ (distinct from $\mathcal{H}'_t$ in Eq. (\ref{eq:Hprime})) and
\begin{eqnarray}
&&\epsilon_A
%=\epsilon_{0, t}+4hs
=
\langle ++--| \mathcal{H}_t | ++--\rangle \nonumber \\
&&
\qquad
=\tilde{J}_{\tilde{z}} \left( -\frac{1}{2} + 2 \rho^2 \right)
+ \nonumber \\
&& \qquad
2 \sqrt{2} \tilde{J}_{\pm\pm} \rho (2 \mu + \nu +\sqrt{1-\mu^2-\nu^2-\rho^2})
\nonumber \\
&& \qquad
-2\tilde{J}_{\pm} \mu (\mu + 2 (\nu + \sqrt{1-\mu^2-\nu^2-\rho^2}))
.
\end{eqnarray}

\begin{figure}
\includegraphics[width=\columnwidth]{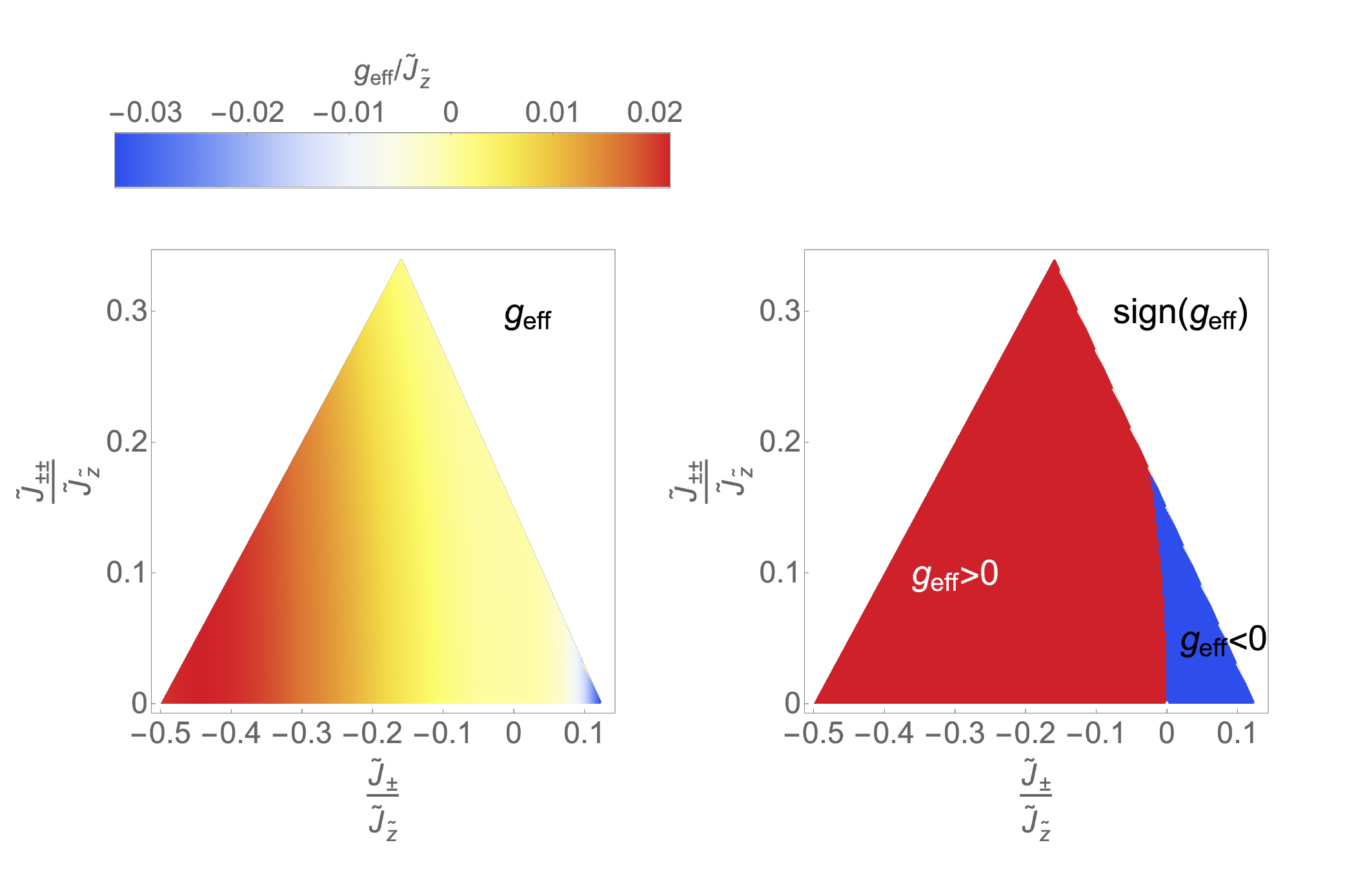}
\caption{Tunnelling matrix element $g_{eff}$, calculated using Eq. (\ref{eq:geff-final}), and the numerically determined CMFT wavefunctions,
plotted over the region of parameter space where the CMFT solutions are highly degenerate.
The sign of $g_{eff}$, shown in the second panel, determines whether a 0-flux or $\pi$-flux U(1) QSL phase is predicted.
}
\label{fig:gplots}
\end{figure}

The contribution of  any `A' tetrahedron which does not change configurations between $ \{ \sigma'\} $
and $ \{ \sigma\} $ to Eq. (\ref{eq:matrix-elements}) vanishes.
Similarly, the contribution of any `B' bond connecting two unchanged `A' tetrahedra vanishes.

There are three kinds of non-vanishing contribution to the matrix element to flip a hexagon.
Firstly, the three `A' tetrahedra belonging to the flipped hexagon (highlighted in red in Fig. \ref{fig:contributions}) contribute:
\begin{eqnarray}
g_A=(o_2)^2 (\eta_2 - o_2 \epsilon_A)
\end{eqnarray}
where
\begin{eqnarray}
&&\eta_2=\langle ++--|\mathcal{H}_t |+-+-\rangle \nonumber \\
&& \qquad
=-\frac{1}{4} \tilde{J}_{\tilde{z}} \left( \mu^2 -6\rho^2+2\mu(\nu+\sqrt{1-\mu^2-\nu^2-\rho^2}) \right)
\nonumber \\
&& \qquad
+ 2 \sqrt{2} \tilde{J}_{\pm\pm} \rho (2 \mu + \nu +\sqrt{1-\mu^2-\nu^2-\rho^2})
 \nonumber\\
&& \qquad
-\tilde{J}_{\pm}  (1-\rho^2+ 2(\mu+\nu) (\mu + \sqrt{1-\mu^2-\nu^2-\rho^2})). \nonumber \\
\end{eqnarray}

Secondly, there are contributions from `B' bonds connecting to the interior of the flipped hexagon 
(highlighted in blue in Fig. \ref{fig:contributions}):
\begin{eqnarray}
g_{B1, ij}=-\tilde{J}_{\tilde{z}}s^2 \sigma_i \sigma_j(o_2)^3.
\end{eqnarray}

Finally, there are contributions from `B' bonds connecting to the exterior of the `A' tetrahedra on the flipped hexagon 
(highlighted in green in Fig. \ref{fig:contributions}):
\begin{eqnarray}
g_{B2, ij}=\tilde{J}_{\tilde{z}}s (\zeta-s) \sigma_i \sigma_j(o_2)^3.
\end{eqnarray}
where
\begin{eqnarray}
&&\zeta=\frac{\langle ++--| \tilde{\tau}_0^{\tilde{z}} |+-+-\rangle}{o_2} \nonumber \\
&&=\frac{\mu \left( -\nu + \sqrt{1-\mu^2-\nu^2-\rho^2} \right)}{\mu^2+2\rho^2+2\mu(\nu+\sqrt{1-\mu^2-\nu^2-\rho^2})}
\nonumber \\
\end{eqnarray}

Summing these contributions and accounting for the fact that $\sigma_i$ must alternate around the hexagon and must
obey Eq. (\ref{eq:icerule}) everywhere, we arrive at the matrix element:
\begin{eqnarray}
g_{eff}=3 (o_2)^2\left(
\eta_2 -\epsilon_A o_2 + \tilde{J}_{\tilde{z}} s^2 o_2 - 2\tilde{J}_{\tilde{z}} s (\zeta-s) o_2
\right) \nonumber \\
\label{eq:geff-final}
\end{eqnarray}

From Eq. (\ref{eq:geff-final}) and the numerically determined CMFT wavefunctions we can calculate $g_{eff}$ and thus predict the ground state
in the degenerate region of CMFT from the sign of $g_{eff}$.
The behavior of $g_{eff}$ and ${\rm sign}(g_{eff})$ over the relevant region of parameter space is shown in Fig. \ref{fig:gplots}.

\bibliography{DOpyrochlores.bib}
%%%%%%%%%%%%%%%%%%%%%%%%%%%%%%%%%%%%%%%%%%%%%%
\end{document}